\documentclass[aps,pre,twocolumn,superscriptaddress,showpacs]{revtex4}
\usepackage{epsfig,amsmath,amsfonts,amssymb,graphics,color,calc}

\newcommand{\be}{\begin{equation}}
\newcommand{\ee}{\end{equation}}
\newcommand{\ba}{\begin{eqnarray}}
\newcommand{\ea}{\end{eqnarray}}

\begin{document}

\title{Finite size effects in the dynamics of glass-forming liquids}

\author{Ludovic Berthier}
\affiliation{Laboratoire Charles Coulomb, UMR 5221, CNRS and 
Universit\'e Montpellier 2, Montpellier, France}

\author{Giulio Biroli}
\affiliation{Institut de Physique 
Th\'eorique (IPhT), CEA, and CNRS URA 2306, 91191 
Gif-sur-Yvette, France}

\author{Daniele Coslovich}
\affiliation{Laboratoire Charles Coulomb, UMR 5221, CNRS and 
Universit\'e Montpellier 2, Montpellier, France}

\author{Walter Kob}
\affiliation{Laboratoire Charles Coulomb, UMR 5221, CNRS and 
Universit\'e Montpellier 2, Montpellier, France}

\author{Cristina Toninelli}
\affiliation{Laboratoire de probabilit\'es et mod\`eles al\'eatoires, 
UMR CNRS 7599, Universit\'e Pierre et Marie Curie et Universit\'e
Denis Diderot, 4 Place Jussieu, 75252 Paris Cedex 05, France}

\date{\today}

\begin{abstract}
We present a comprehensive theoretical study of finite size effects 
in the relaxation dynamics of glass-forming liquids. 
Our analysis is motivated by recent theoretical progress regarding 
the understanding of relevant correlation length scales
in liquids approaching the glass transition. 
We obtain predictions both from general theoretical arguments 
and from a variety of specific perspectives: 
mode-coupling theory, kinetically constrained and defect models,  
and random first order transition theory.
In the latter approach, we predict in particular a non-monotonic evolution  of 
finite size effects  across the mode-coupling crossover due to the competition 
between mode-coupling and activated relaxation.
We study the role of competing relaxation mechanisms in 
giving rise to 
non-monotonic finite size effects by devising a 
kinetically constrained  model where the proximity to the 
mode-coupling singularity can be continuously tuned by changing 
the lattice topology. 
We use our theoretical findings to interpret the results 
of extensive molecular dynamics studies of four model liquids
with distinct structures and kinetic fragilities. While the
less fragile model only displays modest finite size effects, 
we find a more significant size dependence evolving 
with temperature for more fragile models, such as 
Lennard-Jones particles and soft spheres. 
Finally, for a binary mixture of harmonic spheres
we observe the predicted non-monotonic temperature evolution
of finite size effects near the fitted mode-coupling singularity, suggesting 
that the crossover from mode-coupling to activated dynamics is more
pronounced for this model.
Finally, we discuss the close connection between our results  
and the recent report of a non-monotonic temperature evolution 
of a dynamic length scale near the mode-coupling crossover in 
harmonic spheres. 
\end{abstract}


\maketitle

\section{Introduction}

Theoretical studies of the glass transition make heavy use 
of computer simulations of both simple model systems 
such as lattice glass models or kinetically constrained  
spin models, and more realistic models of liquids studied 
through molecular dynamics simulations~\cite{walterhouches}.
Usually, numerical studies are performed  
with periodic boundary conditions using
system sizes that are `large enough'
to provide results that are representative of 
the thermodynamic limit~\cite{allen}.
From this point of view the existence of finite size effects is a
nuisance.
However, as discovered in the context of 
standard critical phenomena~\cite{fssbook},
a thorough study of finite size effects can be very informative: 
It allows one 
to measure the growing correlation lengths, 
to ascertain the critical properties, to 
obtain quantitative information about fluctuations of the order parameter, 
and to provide crucial tests for theoretical approaches. 
While a large amount of work 
has been devoted to measuring the spatial extent
of growing correlation length scales~\cite{book} in supercooled liquids, 
only few studies 
have paid specific attention to finite size 
effects~\cite{jackle1,binder,silica,doliwaheuer,heuer1,kim,jackle2,heuer}, 
while an even smaller number of studies have made 
explicit use of finite size scaling techniques to explore glass 
transitions~\cite{berthierfss,dasguptafss,delgado,proc}.
The aim of this paper is to fill this gap and to address 
from a theoretical point of view 
the issue of finite size scaling in supercooled liquids.   

For quite a long time, the glass transition was considered 
a puzzling phenomenon, corresponding to an obvious change 
between fluid and solid states, but without any of the signs 
found near standard phase transitions, apart from a dramatic
but gradual viscosity increase when approaching the experimental glass 
temperature~\cite{edigernagel,binderkob}. In apparent 
agreement with this situation, early numerical simulations 
did not reveal the strong system size dependences that would 
for instance smear out singular behaviours expected 
near ordinary phase transitions~\cite{ernst,barrat}. 
In more recent years, important progress has been made regarding the status 
of the glassy state and of the relevant length scales 
characterizing systems approaching the glass transition~\cite{rmp}. 
In particular, two decades of active research on dynamic heterogeneity 
in amorphous materials have established 
that the formation of rigid amorphous structures is indeed accompanied by 
nontrivial spatiotemporal fluctuations, which become more pronounced  
upon approaching the 
glassy phase and are characterized by growing dynamic correlation length 
scales~\cite{book}.  

A more recent line of research aims at demonstrating also the existence
of growing static correlation length scales, using point-to-set correlation
functions~\cite{BB,pointtoset,rob1,cavagna,biroli_08,glen}. 
The idea is to confine the system using carefully chosen amorphous
boundary  conditions to detect the existence of 
multi-point static correlations in viscous liquids. 
The point-to-set lengthscale quantifies the spatial extent of 
these correlations. However, 
it is still unclear if such static length scales 
are equivalent to~\cite{cavagna2}, indirectly related to~\cite{rob}, or 
even decoupled from~\cite{sandalo,patrick,PRE12,inprepa}, dynamic ones. 
Actually, the answer to this question may also depend on the level of 
supercooling, thus revealing the existence of 
physically distinct temperature regimes. 

As is well-known in computational studies
of phase transitions, the size of the system can be used as an
additional physical length scale in the problem. Thus, the interplay between 
the system size and the correlation lengths can be used to probe 
the role played by dynamic and static correlation lengths in determining 
the physical behavior of supercooled liquids.
In particular, 
an important  motivation for the present work is the recent 
numerical finding~\cite{sandalo}
that, for a system of harmonic spheres, a surprising 
non-monotonic temperature evolution of dynamic correlation length scales
near an amorphous wall has been detected in the temperature regime 
corresponding to the mode-coupling temperature~\cite{gotze_book}, 
which was interpreted 
as a direct evidence that the physical mechanisms for relaxation 
are different at moderate and low temperatures. If true, 
this interpretation suggests that a similar change of behaviour could 
also occur in the bulk, and could be revealed by
studying carefully finite size effects in the same 
temperature regime. We shall see below that our data indeed 
support the hypothesis of Ref.~\cite{sandalo}, at least 
for the harmonic sphere system.
 
As mentioned above, finite size effects can also be used to test and compare 
theoretical approaches. Indeed, these provide different, and sometimes 
contrasting, 
predictions on the nature and extent of correlation lengths
and fluctuations in viscous liquids~\cite{rmp}. 
For instance kinetically constrained models~\cite{sollich}
and the dynamic facilitation approach~\cite{reviewfacile} focus
on dynamic length scales that usually 
diverge at a zero-temperature dynamic critical 
point~\cite{steve}, random first order transition 
theory~\cite{rfot,bbreviewrfot} predicts
the occurrence of a narrowly avoided mode-coupling 
dynamic singularity in the moderately supercooled
regime, with a crossover towards a 
second regime controlled by a thermodynamic singularity 
and activated dynamics at lower temperatures, each domain
being associated to its own diverging length scale~\cite{silvio07}. 
As a consequence, the interplay between correlation lengths and system size 
and, hence,
the resulting finite size effects depend on the 
theoretical approach. 
A central motivation for the present work is to obtain, discuss and compare to 
numerical simulations 
the theoretical predictions concerning finite size effects. 

We emphasize that since our focus are
highly viscous supercooled liquids, we do not discuss
the literature about finite size effects in simple 
liquids which is a different topic, for which 
hydrodynamic effects, ignored here, play 
a more central role~\cite{bocquet,jackle}.  

In summary, this paper presents  
a comprehensive theoretical study of finite size effects 
in supercooled liquids. Our aim 
is to provide useful practical information about the relevance 
of finite size effects in computer studies of the glass formation, 
test theoretical approaches
and also to obtain new insights about the nature of the fluctuations 
revealed by finite size studies, in particular in the region
of the mode-coupling crossover.

The paper is organized as follows. In 
Sec.~\ref{theory1} we provide general theoretical arguments 
about confined supercooled liquids. In 
Sec.~\ref{theory2}, we use specific theoretical approaches 
to make predictions regarding finite size effects. 
In Sec.~\ref{rfa} we introduce a new lattice glass model 
with an avoided  mode-coupling singularity whose strength can be tuned 
by changing the lattice topology and use it to study 
finite size effects.
In Sec.~\ref{md} we provide molecular dynamics simulations 
of four model liquids with distinct structures and kinetic 
fragilities. In Sec.~\ref{conclusion} we close the paper with 
a discussion of the results. 
More details about the new lattice glass model introduced 
in Sec.~\ref{rfa} are given in the Appendix.

\section{Emergence of finite size effects in fragile liquids}

\label{theory1}

The relaxation time of glass-forming materials
approaching the glass transition follows a thermally activated 
form,  
\be
\tau_\alpha \approx \tau_0 \exp \left[ \frac{E(T)}{T} \right],
\label{superarrh}
\ee
where the activation energy $E(T)$ increases when temperature 
decreases for fragile materials, whereas it is constant 
for strong glass-formers~\cite{rmp}. Here and in the following 
we shall absorb the Boltzmann constant $k_B$ in the definition of $T$. 
The temperature dependence in Eq.~(\ref{superarrh})
means that the nature of the relaxing `entities', 
whatever they are, 
changes with temperature. A growing activation energy 
actually suggests an increasing cooperativity in the relaxation events, 
corresponding 
to the correlated motion of an increasing number of particles. Thus, by 
reducing the system 
size one expects, all other things being equal, that the dynamics 
starts to differ from bulk behavior
when the linear size becomes comparable to the size of these correlated 
regions.  Thus, it is
natural to expect the emergence of a characteristic length scale, 
which we denote by $\ell_{FS}(T)$ in the following, characterizing 
finite size effects such that bulk relaxation is obtained 
when the system size is larger than $\ell_{FS}(T)$.
In cases where the dynamics is characterized by several 
length scales, the behaviour of $\ell_{FS}(T)$ will be more complicated,
as we shall see.  
In any case, we believe that finite size studies should provide 
a new way to probe dynamical correlations and cooperativity in 
glass-forming liquids~\cite{dasguptafss}.

By using a very general argument, one can show that 
$\ell_{FS}(T)$ must grow for fragile liquids 
when temperature decreases. The starting point of our argument is to recall 
that an upper bound for the relaxation
timescale for a system of linear size $L$ can be obtained assuming the 
worst case scenario, namely that
all particles have to move together in a cooperative way 
to relax the structure. This leads to an upper 
bound for $\tau_\alpha(L,T)$ that scales with $L$ as 
\be
\label{ub}
\tau_\alpha(L,T) \le 
\tau_{ub}(L,T) = \tau_0 \exp\left(\frac{c L^d}{T} \right),
\ee
where $\tau_0 $ is a microscopic timescale, $c$ a 
numerical constant, and $d$ the spatial dimension. 
Note that for systems evolving with stochastic dynamics and with discrete 
degrees of freedom this is a rigorous statement~\cite{cristina}. For 
other systems, this result is expected on general ground but a rigorous proof
is probably out of reach.

Now, consider an infinite system characterized 
by the relaxation timescale $\tau_\alpha(T)=\tau_\alpha(L\to\infty,T)$, 
and then decrease its size while simultaneously measuring  
$\tau_\alpha(L,T)$. By definition,
the structural relaxation time will not change until $\ell_{FS}(T)$ 
is reached. A lower bound for this length can be obtained by noticing that a 
constant $\tau_\alpha(L,T)$ as a function of $L$ would necessarily 
violate the bound in Eq.~(\ref{ub}) at small 
$L$ and large enough $\tau_\alpha(T)$.
Thus, dynamical finite 
size effects must appear when (or before that) $\tau_\alpha(\infty,T)$ becomes 
equal to  $\tau_{ub}(L,T)$. Therefore, we find 
\be 
\ell_{FS}(T) \ge \left( \frac T c \ln \left[ \frac{\tau_\alpha(T)}{\tau_0}
\right]\right)^{1/d} .
\ee
Using Eq.~(\ref{superarrh}),
this result implies that, up to a proportionality constant, $\ell_{FS}(T)$ must 
increase with temperature at least as $E(T)^{1/d}$. 
Note that this result does not imply anything 
about the precise dependence of 
$\tau_\alpha$ on $L$, in particular whether this 
dependence is monotonic or not.  
However, it shows that for fragile liquids
the length obtained by dynamical finite size effects 
studies has to increase when temperature decreases. 
This growth would be  
faster than $(1/T)^{1/d}$ or $(T/(T-T_{VFT}))^{1/d}$, in the 
respective cases where 
$\tau_\alpha(T)$ follows a B\"assler or a Vogel-Fulcher temperature
dependence. 

At this point, three important remarks are in order. 

(1) The lower bound on $\ell_{FS}(T)$ only becomes meaningful when 
$(T\ln [\tau_\alpha(T)/\tau_0]/c)^{1/d}\ge a$, where $a$ is the typical 
interparticle distance. The temperature at which this takes place 
of course depends 
on the values of the constant $c$ and of $\tau_0$ and may actually correspond 
to very deep supercooling, i.e. very large values of $\tau_\alpha$. It 
would be interesting to have an estimate of $c$ and $\tau_0$ for a given
liquid to understand what is the highest temperature at which our argument 
becomes useful.

(2) For a strong (i.e. Arrhenius) 
liquid, $E(T)=E$, one finds that $(T\ln [\tau_\alpha(T)/
\tau_0]/c)^{1/d}$ does not depend on temperature. This makes perfect 
sense within the physical picture where strong liquids 
relax by localized and independent thermally activated events.
In this case $\ell_{FS}$ should be temperature independent and 
roughly equal or at least proportional 
to the microscopic length scale $a$.

(3)
The lower bound we obtained for $\ell_{FS}(T)$ actually coincides with 
the one obtained in Ref.~\cite{pointtoset} 
for the static point-to-set length. This is reasonable because one 
expects the dynamical length probed by finite size effects to be 
larger than (or equal to) the static point-to-set length.  

\section{Specific theoretical predictions}

\label{theory2}

Having argued by general arguments that $\ell_{FS}(T)$ increases 
when $T$ decreases for fragile liquids and hence should be related to 
cooperative relaxation of some sort, we now address the precise 
form of $\tau_\alpha(L,T)$ as a function 
of $L$ and the possible physical mechanisms behind the increase of 
$\ell_{FS}(T)$. Since this partially depends on the 
particular theoretical description used, we consider several 
different cases below. 

An important conclusion of the following sections is that 
$\ell_{FS}(T)$ cannot be univocally related to one of the several 
correlation lengths introduced recently. 
This relation depends on the theory: $\ell_{FS}(T)$ may coincide with 
a static correlation, like the point-to-set one, or with the dynamic 
correlation length, or only indirectly related 
to either one of them.   

\subsection{Cooperatively rearranging regions}
\label{crr}

There are several theories that explain the relaxation process
in supercooled liquids in terms of  
cooperative rearrangements of regions involving a growing number of 
particles.  
Theories falling in this category are the Adam-Gibbs theory~\cite{adam_65}, 
the 
frustration-limited domain approach~\cite{gillesphysica,gillesreview} 
and Random First Order Transition Theory 
(RFOT)~\cite{rfot,rfotwolynes,bbreviewrfot}. 
For the latter, a different dynamical process 
described by mode-coupling  theory~\cite{gotze_book} 
is responsible for relaxation for temperatures larger than $T_{MCT}$,
and this regime is discussed separately below in Sec.~\ref{mct}.
 
In all these theories the relaxation time is derived by assuming activated 
dynamic scaling and using 
as a characteristic length scale the linear size of the rearranging regions. 
Thus, the logarithm of the 
relaxation timescale is proportional to the length scale raised 
to some power, which for instance is equal to $d$ 
in the Adam-Gibbs case~\cite{adam_65}. The physical mechanism 
responsible for the growth of this length scale and the exponent of the 
power law 
depend on the details of the theory. 

Assuming that cooperative relaxation events are 
uncorrelated,
as it is usually done, the primary effect of decreasing system size  
is to decrease the value 
of the activation energy from its bulk value, because the number of particles 
involved decreases, once the 
system size becomes smaller than the typical size of a 
rearranging region.  
Thus, for all these theories, one expects $\tau_{\alpha}(L,T)$ to be a 
monotonically increasing function of $L$ approaching the bulk value 
from below for 
$L$ of the order of the size of the rearranging regions. 
This is directly reminiscent of finite size effects near 
a second order phase transition~\cite{fssbook}, where 
divergences are smeared out by finite system sizes. 
The only peculiarity of the glass transition would be the occurence
of activated, rather than algebraic, forms of dynamic scaling. 
It would be interesting to know whether some kind of 
scaling formula holds for $\tau_\alpha(L,T)$ (or $\log \tau_\alpha(L,T)$).

The behavior of $\tau_{\alpha}(L,T)$ as a function of $L$ is also similar to
the one expected within RFOT using amorphous boundary 
conditions \cite{rmp,cavagna2}.
However, the physical mechanisms conjectured to play a role for small 
$L$ are different from the periodic boundary conditions 
considered here. 
In the latter, the dynamics accelerates because the 
boundary condition lowers the free energy of a single state. Therefore  
relaxation occurs inside this state and is hence faster than in the activated 
regime in which the other states should also be visited. 
With periodic boundary conditions 
instead, the dynamics is activated but the barrier 
decreases if $L$ becomes smaller. 
 
\subsection{Defect models}
\label{kcm}

Defect models~\cite{sollich}, and in particular dynamical facilitation 
models~\cite{reviewfacile}, 
have been widely used to study dynamical heterogeneities and spatial 
correlations in viscous liquids, but dynamical finite size effects 
have not been specifically discussed.

In defect models, whether cooperative or not~\cite{sollich}, there are 
at least two relevant length scales. One 
is the typical distance between defects, $\xi_d\propto c^{-1/d}$
($c$ is the defect density), and the 
other one is related to the size of 
dynamical correlations. The two are not 
necessarily equal, and the latter can possibly be much 
larger than the former depending
of the model and the dimensionality~\cite{TWBBB}. Assuming that 
the equilibrium concentration of defects is unaffected by confinement, 
we expect $\xi_d$ to be the most relevant length scale
for the finite size effects.
In fact, for intermediate system sizes  (i.e. for $L$ 
between $\xi_d$ and 
the dynamic correlation
length), one might expect that $\tau(L,T)$ deviates 
only weakly from its bulk value, as the nature of dynamical facilitation
remains essentially unaffected. This statement is straightforward
for diffusing defects and non-cooperative constrained models,
but has to be taken with some caution for cooperative models
which have a more complicated dynamics typically 
characterized by several (and possibly a hierarchy) 
dynamic length scales~\cite{jackle1,sollich}.  

By contrast, the nature of the dynamical processes must change
qualitatively when $L$ competes with $\xi_d$. 
Indeed, for $L \simeq \xi_d$ about half of the equilibrium configurations 
contain strictly no defect. In this case, the system has to either use a 
different
channel for relaxation or create a new
defect and then relax by defect diffusion. 
(Note that this second scenario is strictly forbidden in 
spin facilitated models, 
which thus become instead non-ergodic in this limit~\cite{sollich}.) 
In both cases
the corresponding relaxation time is expected to be larger than the one for 
configurations 
having a defect from the beginning, which instead relax on a timescale of the 
order of the 
bulk relaxation time. Therefore, 
the average relaxation time is expected 
to start to increase strongly when  
$L \simeq \xi_d$ (and to become infinite
in constrained spin models). 
The behavior for $L<\xi_d$ is less clear because in this 
case the system 
typically does not have any defect in equilibrium configurations and, hence, 
relaxes in 
a way different from the one used for $L \gg \xi_d$, and no alternative 
relaxation channel is described within the defect approach.
Since in this case the shape of $\tau_{\alpha}(L,T)$
is determined by the $L$ dependence of this unknown relaxation mechanism 
we cannot say much about it. 
However, using the general argument developed above we 
know that 
for fragile liquids  $\tau_{\alpha}(L,T)$ has to decrease when the system 
size is reduced below 
$(T\ln [\tau_\alpha(T)/\tau_0]/c)^{1/d}$. Thus, in the case of fragile 
liquids and within the framework of defect
diffusion theories we expect $\tau_{\alpha}(L,T)$ to 
have a non-monotonic behavior 
that can be more or less pronounced depending on the underlying 
model. Instead, in the case of 
Arrhenius liquids, it is possible that 
$\tau_{\alpha}(L,T)$ increases with decreasing $L$ until the linear 
system size is almost of the order of the interparticle distance.  

As final comments, we first stress that some 
form of facilitation dynamics could additionnally be introduced  
within cooperative rearranging regions theories. 
Thus some mild non-monotonous behavior 
can be present even in that case. Second, whatever is the correct theory, 
there could be actually several length scales playing a role. It is likely 
that larger length scales
are affected first by the confinement such that very large domains
disappear first when $L$ is reduced, which should somehow truncate 
the distribution of relaxation times, making the average (e.g the first 
moment of the distribution) smaller. This suggests that finite 
size effects could manifest themselves at very long times only,
which correspond to the final decay of time correlation functions. 

\subsection{The mode-coupling theory crossover}
\label{mct}

In the framework of the mode-coupling theory of the glass 
transition~\cite{gotze_book},
a dynamical transition
accompanied by a diverging dynamic correlation length scale~\cite{BBEPL} 
takes place at 
a finite temperature $T_{MCT}$. It is well-known that the transition 
predicted by the theory actually 
does not take place in finite dimensional models and in experiments.
Instead, it is replaced by a crossover occurring 
near the temperature extracted from fitting the dynamical relaxation 
to scaling predictions of the theory, although the precise nature 
of this crossover is not well understood~\cite{xover,bbreviewrfot}.
RFOT theory naturally includes mode-coupling theory, 
but at present it cannot describe the nature of the crossover from
mode-coupling to activated relaxation in much detail~\cite{bbreviewrfot}.  
Decreasing $L$ at constant temperature above 
$T_{MCT}$ (and, of course, below the onset temperature), 
we again expect a non-monotonic size dependence 
of $\tau_{\alpha}(L,T)$  due to the competition between
two effects, which we now explain.

When considered from the point of view of 
random first order transitions~\cite{bbreviewrfot}, a physical interpretation 
of the dynamics within the mode-coupling regime 
is that relaxation occurs 
along unstable modes that become less and less unstable
and more collective upon approaching $T_{MCT}$, thus
leading to diverging dynamic correlations~\cite{BBEPL}.
By reducing the system size below the dynamic correlation length, 
these unstable 
extended modes are the most easily affected, 
and confinement should render some of these modes stable,
thus closing some relaxation channels. The effect would then be
to slow down relaxation, or equivalently to shift the apparent 
value of $T_{MCT}$ to larger temperature~\cite{silviobook,silvio07}. 

However, decreasing the system size has a second consequence.
The singularity predicted by the theory cannot occur in a finite 
size system, and therefore   
$\tau_{\alpha}(L,T)$ cannot increase indefinitely by decreasing $L$. 
In fact this growth is bounded from above by the 
general argument detailed in Sec.~\ref{theory1}
even though this may happen on quite a small 
length scale. Within
RFOT theory, the 
increase of $\tau_{\alpha}(L,T)$ at small $L$ should 
be cut off for length scales
such that relaxing via the mode-coupling channel becomes 
slower than the activation channel. 
Then, for even smaller $L$, relaxation will proceed by cooperative 
rearrangements (as it does in the bulk below $T_{MCT}$) 
and this should lead, as in Sec.~\ref{crr}, 
to a decrease of  $\tau_{\alpha}(L,T)$ by decreasing $L$.

Overall, we should then observe a non-monotonic behaviour 
of the relaxation time in the mode-coupling regime, because
using systems with finite sizes has a qualitatively different 
impact on mode-coupling dynamics (which slows down in confinement)
and activated dynamics (which accelerates in confinement).

Although already quite complex, the picture depicted 
above is certainly still too simplistic. As stated before, 
the mode-coupling crossover is not well understood and 
it is thus likely that dynamical finite size effects 
will turn out to be quite subtle and lead to a very complex behaviour, as 
found in seemingly simpler mean-field models~\cite{brangian,sarlat,parisi}.
In order to shed some more light on this crossover, 
we consider in the following a 
finite dimensional model that displays an avoided 
mode-coupling singularity.
 
\section{A lattice glass model with an avoided mode-coupling transition}
\label{rfa}

\subsection{Kac-Fredrickson-Andersen (KFA) model}

We study a two dimensional Kac-version of the spin facilitated model 
extensively studied by Fredrickson and Andersen~\cite{FA,FAb,FA2},
which we call the `Kac-Fredrickson-Andersen' (KFA) model.
This is defined by a non-interacting Hamiltonian
\be
H = \sum_i n_i,
\label{FA}
\ee
where $n_i = 0, 1$ represents a binary mobility defect variable.
The average density of spins in the excited state is $c(T) = 
(1+\exp(1/T))^{-1}$.
For the dynamics, we choose the 2-spin facilitation rule, as in the original
model~\cite{FA}: 
To be able to flip, a spin must possess at least 2 neigbours which are 
both in the excited state $n_i=1$. 

A generalisation to a Kac-version of the model can be  
obtained by introducing a new geometrical parameter, 
$K$, which characterizes the range of the spin connectivity within 
a regular two-dimensional square lattice. 
In the standard FA model in two dimensions, 
the spin at position ${\bf r}$ has 4
nearest neighbours 
that contribute to facilitating its dynamics.
They occupy the positions 
${\bf r} +  {\bf e_x}$,
${\bf r} -  {\bf e_x}$,
${\bf r} +  {\bf e_y}$, and
${\bf r} -  {\bf e_y}$,
where ${\bf e_x}$ and ${\bf e_y}$ are unit vectors along the horizontal
and vertical directions, respectively.
In our Kac-version, the spin at position ${\bf r}$ remains 
facilitated by 4 `neighbors' which are now 
located at positions 
${\bf r} +  i {\bf e_x}$,
${\bf r} -  j {\bf e_x}$,
${\bf r} +  k {\bf e_y}$, and
${\bf r} -  l {\bf e_y}$,
where $(i, j, k, l)$ are 4 random numbers 
chosen in the set $\{1, \cdots, K\}$  
according to a procedure that will be detailed below.
Thus, the KFA model contains quenched disorder, and 
spins can interact on the lattice over a range $K$ that can be tuned
at will and be made arbitrarily large, while keeping the spin connectivity
constant and equal to that of the original $K=1$ model. 

To generate an instance of the lattice, we start from the leftmost site of
a given line and connect it at random to one of its $K$ right-neighbors. We
then move to the right by one lattice spacing and connect the new site at
random to one of its $K$ right-neighbors, excluding the one that has been
connected at the previous step. This procedure is then iterated up to the
$(K-1)$-th site of the line always connecting the current site at random to one
of its $K$ right neighbors excluding the ones that have been previously
connected. In order to guarantee that at the end of the procedure each site
has a right and a left neighbour we proceed as follows for the remaining
sites. When choosing the right neighbour of site $i$ we first check whether
$i+1$ already has a left neighbour. If this is not the case we connect $i$ to
$i+1$, otherwise we pick the right neighbor of $i$ uniformly at random
among the sites $i+2,\dots,i+K$ which have not yet been assigned a left
neighbour. We then iterate this procedure up to the end of the line imposing
periodic boundary conditions. In this way we generate a random one
dimensional lattice with connectivity two at each site and a range equal to
$K$. We then repeat the procedure for each line and column of the square 
lattice to define an instance of the KFA model. When performing simulations 
of the KFA model, we also average over independent realizations 
(typically 2000) of the lattice.

Note that in the limit $K \to \infty$ the probability $\epsilon$ that a site
is connected to its right nearest neighbour does not decay to zero; however
it can be easily be bounded from above by $\exp(-1)$ and the numerics
indicates that this probability saturates at a lower value
$\epsilon \simeq \exp(-2)$.  Thus in the $K \to \infty$ limit,  
the geometry becomes the one of a Bethe lattice  
with connectivity equal to four decorated by loops
occurring with probability at most $\epsilon^4 \simeq \exp(-8)$
and a finite temperature singularity occurs. We expect this transition to
shares important similarities with the mode-coupling transition as it is the
case on the pure Bethe lattice~\cite{MS}. In particular, time correlation
functions should decay in a two-step manner with power laws characterizing
the approach and departure from the intermediate time plateau, and an
algebraic divergence of the temperature evolution of the equilibrium
relaxation time. 
Instead for $K=1$ the topology of the square lattice is recovered and
the physical behavior is the one of a cooperative kinetically constrained
model, with a divergence of timescales and length scales at $T = 0$ 
only~\cite{TBF}.

The crucial point for our purposes is that, for a large 
but finite $K$, we expect the dynamics to be controlled 
in some temperature range by the mode-coupling singularity 
because the system locally resembles the $K=\infty$ decorated tree, 
but the singularity must be avoided 
because the square lattice topology eventually dominates at large enough
length scales. 

Therefore, as is believed to be the case in finite 
dimensional glasses, the mode-coupling singularity is `avoided'
when temperature decreases because of the presence of 
some `activated processes'. 
Actually, in this case it is more correct to call these process 
cooperative instead 
of activated since they are due to the diffusion of macro-vacancies \cite{TBF}.
There are however two important differences 
between the KFA model and supercooled liquids.
First, in the KFA model both temperature regimes and the related 
physical behaviours are well understood, and thus our model 
smoothly interpolates between limits that are under control.
Second, we can tune the parameter $K$ to be very 
close or very far from the mode-coupling limit, and thus we can 
control the relevance of the infinite range dynamics 
for the finite range model, which is not readily realized 
in liquids. We note that a closely related attempt to control the importance 
of mean-field behaviour has recently been
published~\cite{romain} in the context of 
off-lattice models of liquids, 
following an idea similar to ours where the connectivity 
is kept constant while increasing the range of the interaction
between particles. 

In the following we present results of extensive Monte-Carlo simulations 
of the above model using periodic 
boundary conditions and a lattice of linear size $L$ which we vary.
We use a continuous time Monte-Carlo algorithm~\cite{newman}, and study
a broad range of parameters, changing in particular 
the Kac-range from $K=1$ to $K=24$ (note that we only consider the regime 
$K < L/2$). It is important to stress 
that depending on $K$ and $L$, especially at small system sizes, 
some samples may contain a finite fraction of blocked spins,
a `backbone', that will never flip. We disregard these configurations
and sample only the ones that do not contain blocked backbones.
A different choice would be to average over all 
configurations and estimate the relaxation time as the time decay 
of the persistence to a non-zero long-time value. 
We have chosen the first solution, and so we perform a `biased' 
average over all ergodic configurations, in order not to mix 
ergodic and non-ergodic configurations in a single average.
By contrast, no blocked configurations are found when $L$ is large
enough and `bulk' dynamics is studied. 
Here, $L$ `large enough' means that the relaxation time 
and the correlation functions `do not depend on $L$ any more' 
(the size dependence of these functions is 
studied in great detail below).  Because the Hamiltonian 
in Eq.~(\ref{FA}) is trivial, it is straightforward to generate
equilibrium configurations and to study equilibrium 
relaxation in the absence of any aging effect. 
The only limitation is that complete relaxation 
cannot be observed over our finite time window when temperature 
is too low, a regime which again depends on the value of $K$,
and is of course set by our computer resources.  
 
\subsection{Evidence of a mode-coupling crossover for the 
bulk dynamics}

Following previous work~\cite{steve}, we first probe the bulk equilibrium 
dynamics in the KFA model by measuring the persistence function, 
\be 
P(t) =
\left\langle \frac{1}{N} \sum_{i=1}^N P_i(t) \right\rangle,
\ee 
where $P_i(t) =1$ if the spin at site $i$ has not changed state 
between times $0$ and $t$, and $P_i(t)=0$ otherwise.
We have also studied the spin-spin autocorrelation function, 
and have found qualitatively similar results, which are thus 
not presented here. 

\begin{figure}
\psfig{file=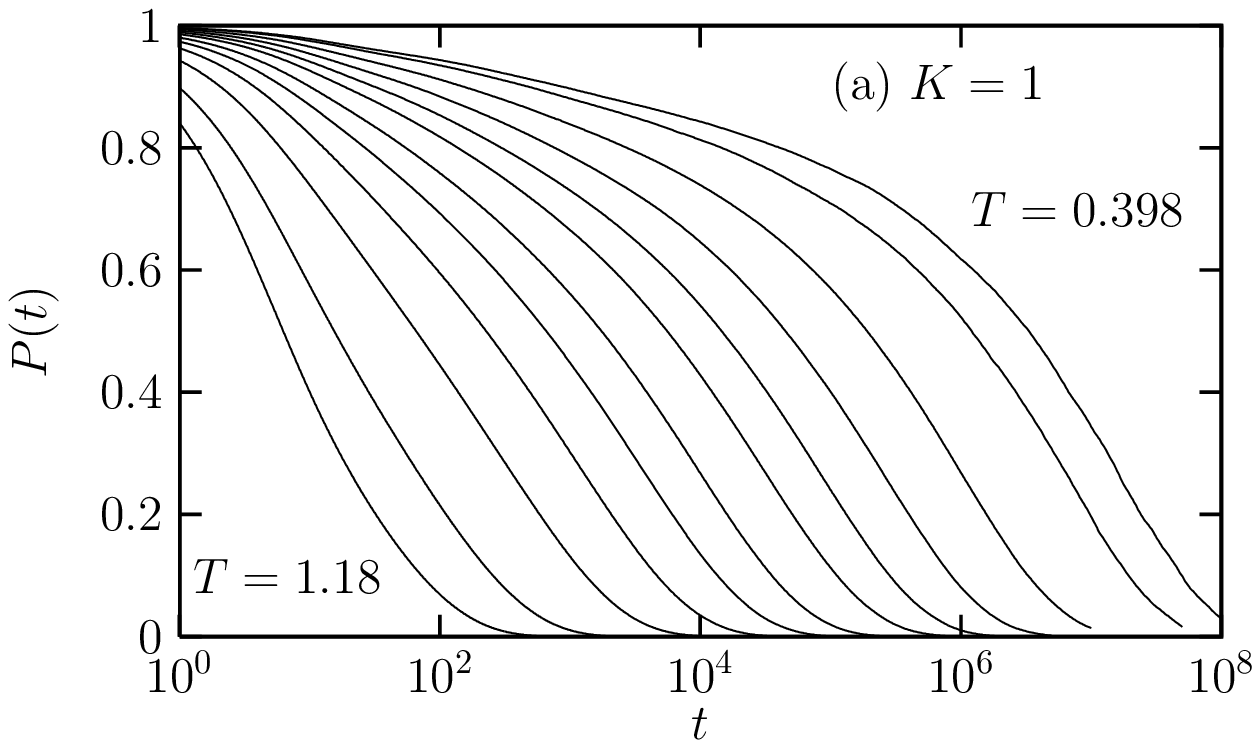,width=8.5cm}
\psfig{file=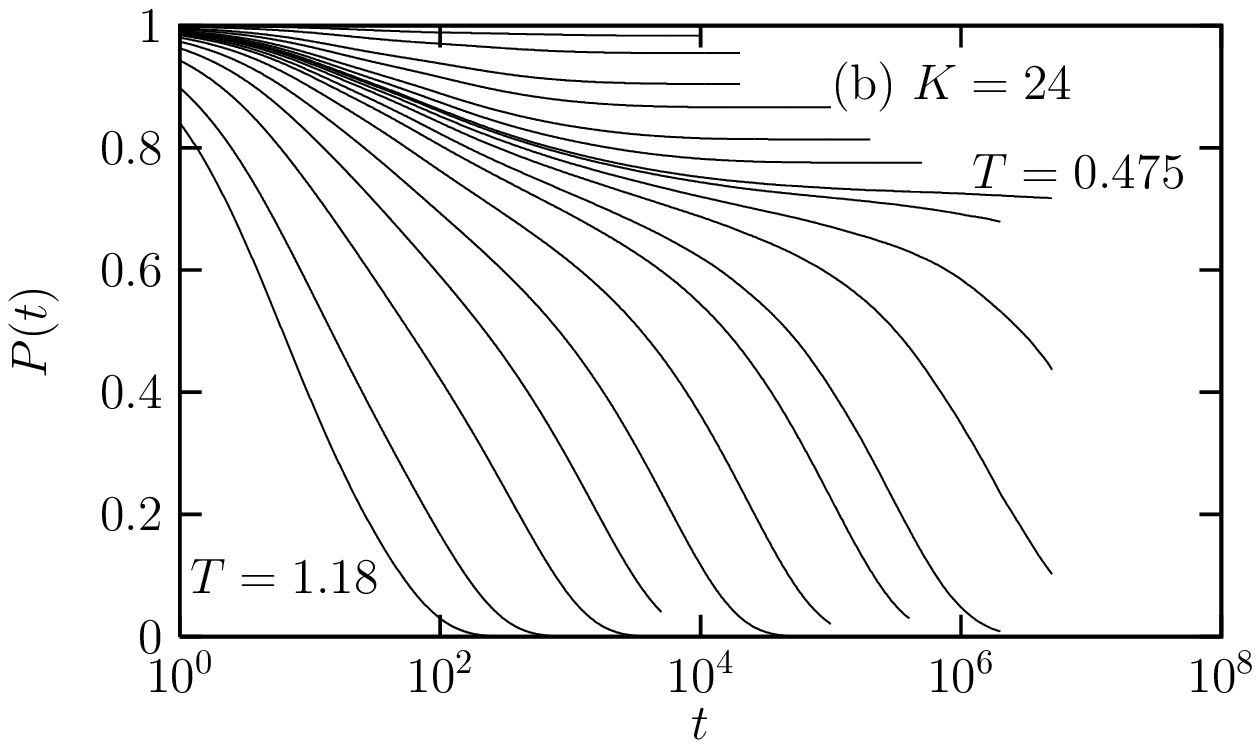,width=8.5cm}
\psfig{file=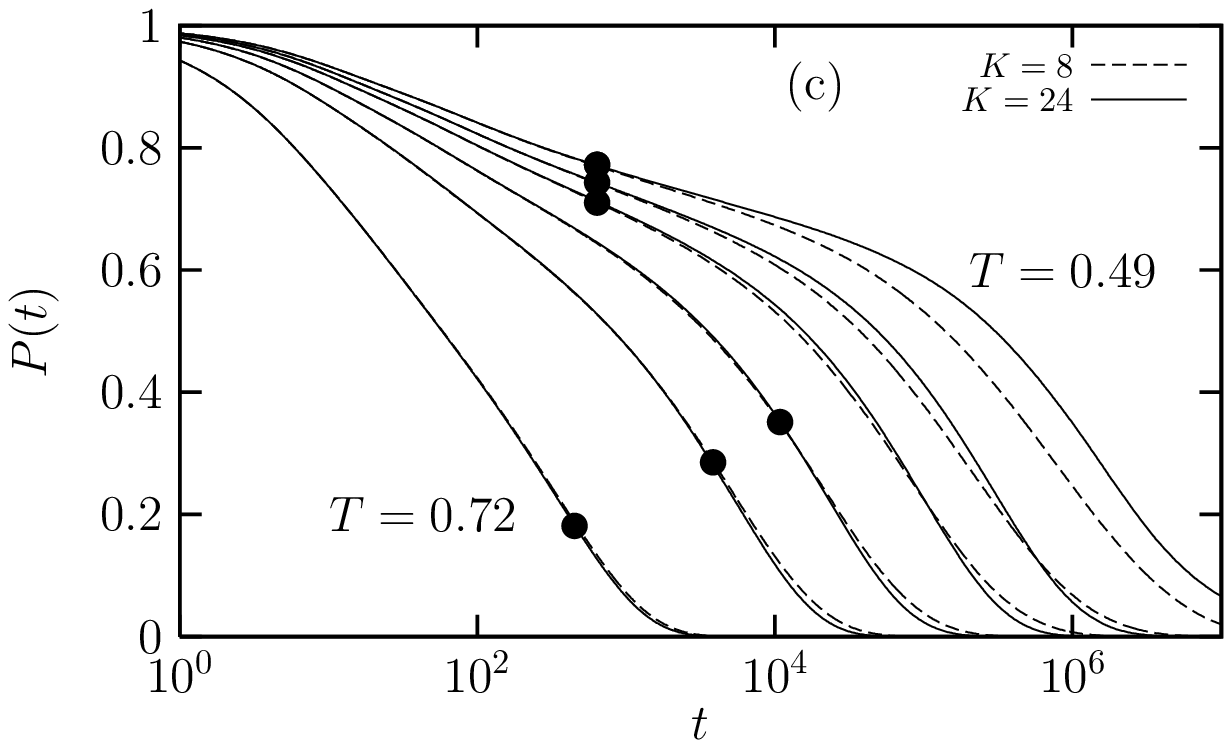,width=8.5cm}
\caption{\label{pers} Time dependence 
of persistence functions in the KFA model
for various connectivity range $K$
and temperatures $T$.
(a) Data for $K=1$ and decreasing temperatures (from left to right). 
(b) The data for $K=24$ and decreasing temperatures  (from 
left to right) clearly show a two-step decay towards an 
intermediate plateau.
(c) Comparison between  data for $K=8$ and $K=24$ at various 
temperatures. A filled circle indicates the crossover timescale
$t^*$ after which the two sets of data deviate significantly, 
thus delimiting the $(t,T)$ domain of applicability of mean-field 
dynamics.}
\end{figure}

Our findings for different connectivity ranges $K$, and 
temperatures $T$ are presented in Fig.~\ref{pers}.
To obtain these data, we have used $L=150$ for $K=1$, up to 
$L=250$ for $K=24$, carefully checking the absence of 
finite size effects. 
The data for $K=1$ in Fig.~\ref{pers}a
resemble previously published results 
in cooperative kinetically constrained 
models~\cite{sollich,steve,KA,MarcoT,nef}. 
The shape of the 
persistence function changes very little when temperature 
decreases. An important point for us is that these 
curves only display a single decay towards zero instead of the 
clear two-step decay (termed alpha and beta relaxations) 
typically found in supercooled liquids.
This distinction was often interpreted as being a consequence
of working on the lattice because short-time thermal vibrations 
cannot contribute to relaxation functions~\cite{KA}.
It was later understood that an analog of the beta-relaxation
could nevertheless appear and be studied in lattice 
models~\cite{MarcoT,MS,nef,xover}, as we now confirm.

When moving from $K=1$ to $K=24$ in Fig.~\ref{pers}b, which is 
the largest range studied in this work,
we find that the shape of the correlators changes 
continuously with increasing $K$. In particular, the short-time 
dynamics changes from being convex (or quasi-logarithmic~\cite{nef,KA})
to becoming concave and converging to an intermediate-time 
plateau, as can clearly be seen in Fig.~\ref{pers}b.  
This two-step decay is also characteristic of the decay
of dynamic correlators predicted by mode-coupling theory.
Due to the small probability ($\simeq \exp(-8)$) for a site to be in a loop,
we expect the system for $K\to\infty$ to be well described by a 
Bethe lattice for the system sizes which we can numerically explore.
We have performed a few simulations directly on the Bethe lattice,
as in Ref.~\cite{MS}, and found that, on the range 
of temperature accessible for $K=24$, the relaxation functions for
$K=24$ and $K=\infty$ are extremely close, and so we take $K=24$ as being 
representative of the infinite-range limit over the accessible 
temperature range. For the studied connectivity, 
the transition temperature for the Bethe lattice limit is 
$T_c = 0.480898$.

Overall, these curves suggest that, as announced,
the KFA model crosses over from a mode-coupling like 
dynamics at large $K$ towards a dynamics of a different nature
at small $K$, which is cooperative, 
yielding a super-Arrhenius growth of the relaxation time.

Interestingly, while the data for $K=1$ appear qualitatively 
different from the $K=\infty$ counterpart, we find that 
for intermediate $K$ values, a finite temperature regime seems to open
where the dynamics is qualitatively similar to the mean-field regime, 
with deviations only appearing at lower temperatures. 
We confirm these observations in Fig.~\ref{pers}c where 
relaxation data for $K=8$ and $K=24$ 
are superimposed for various temperatures. It is clear 
that deviations between both sets of curves are very small 
at high temperatures, and become quite large when $T$ decreases. 
More precisely, we observe that for each temperature the
relaxation data are very similar at short times, but 
differ at long times. This allows the definition of a crossover 
timescale $t^*(T)$, marked with a closed symbol in 
Fig.~\ref{pers}c, such that differences between 
the two persistence functions only become significant 
for $t>t^*$. We find that $t^*$ belongs to the alpha-relaxation
when temperature is not too low, which implies that the 
correlation functions and thus the relaxation
time are controlled, in this temperature regime, by the infinite-range
dynamics. However, when temperature is decreased further, 
$t^*$ now belongs to the beta-relaxation. This means 
that sufficiently close to the dynamic singularity
of the $K=\infty$ model, the temperature evolution of 
$\tau_\alpha$ differs significantly from the infinite-range model, 
so that the singularity is eventually avoided. Quite remarkably, 
we also find that the beta-relaxation does not seem 
to be very much affected by this crossover, which could imply that
short-time dynamics remains well-described by the mean-field 
limit even at temperatures where the long-time dynamics
is already controlled by actived processes. 
This would suggest that a more precise prescription 
for the applicability of the mode-coupling predictions 
should be done in the (time, temperature) domain, rather than
by defining a single crossover temperature~\cite{bbreviewrfot}.
In practice, this suggests that mode-coupling theory could 
still be useful below $T_{MCT}$, at least to describe the short-time
dynamics of viscous liquids and estimate for instance the temperature
evolution of the Debye-Waller factor~\cite{gotze_book,hoping}. 

\begin{figure}
\psfig{file=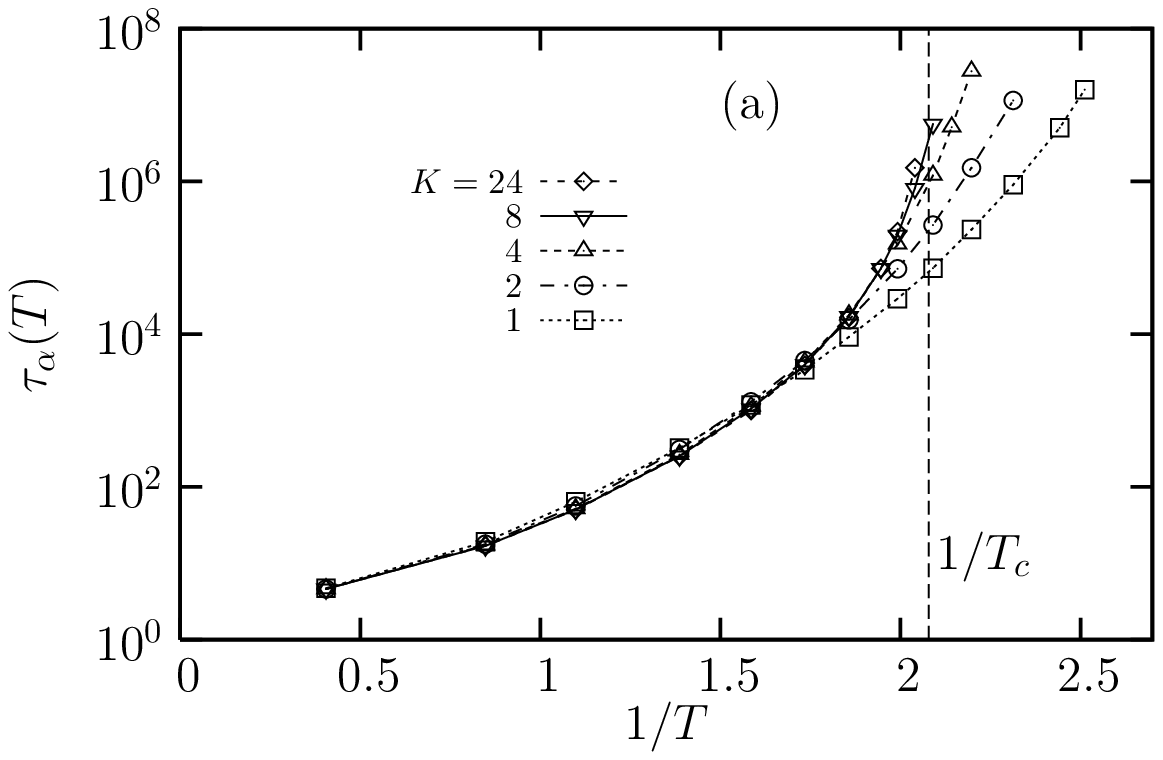,width=8.5cm}
\psfig{file=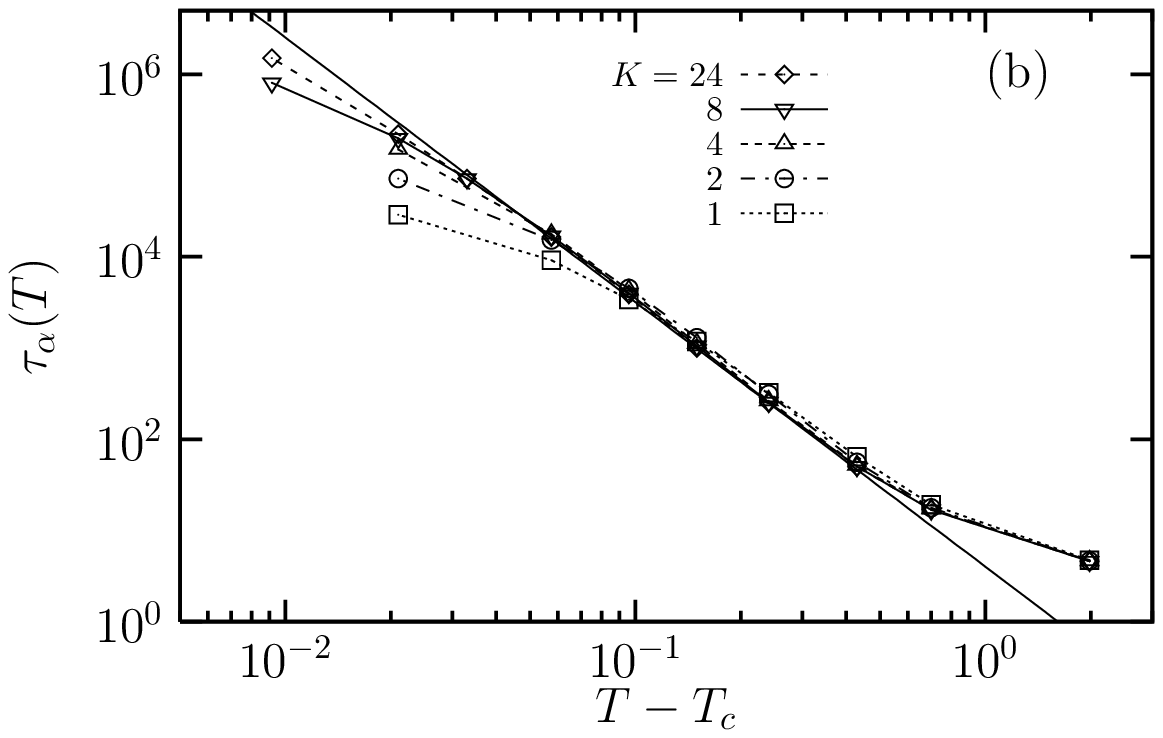,width=8.5cm}
\caption{\label{tau} 
Temperature evolution of the relaxation time 
in the KFA model. (a) Arrhenius plot, the vertical 
line is at $T_c (K=\infty) = 0.480898$.
(b) The same data plotted in a log-log representation as a function of 
$T-T_c$, the full line representing the mean-field limit of 
Eq.~(\ref{powerlaw}). In both panels, we see that deviations 
from mean-field occur at higher $T$ for smaller $K$, which shrinks
the domain of validity of the mean-field prediction for the 
growth of $\tau_\alpha$.} 
\end{figure}

A much sharper identification of the crossover is found by studying the 
alpha-relaxation time of the system. We extract $\tau_\alpha(T)$ 
from the relaxation curve using the
definition $P(\tau_\alpha(T)) = const$. Note that the value of the constant 
is irrelevant as the shape of the relaxation 
barely evolves with temperature for a given value of $K$, as long as 
the constant corresponds to the final decay. We use 
the value 0.28 which lies roughly in the middle of the final 
relaxation for large $K$. 
The temperature dependence of the corresponding 
$\tau_{\alpha}(T)$ is presented in Fig.~\ref{tau} 
using an Arrhenius plot, see Fig.~\ref{tau}a, or using a reduced 
plot inspired by the mode-coupling prediction, see Fig.~\ref{tau}b.

These results confirm that for the largest range studied, $K=24$,
the dynamics is very close to the results obtained directly on 
the Bethe lattice~\cite{MS}, which were shown to obey a power 
law divergence, 
\be 
\tau_{\alpha}(T) \propto |T-T_c|^{-\gamma},
\label{powerlaw}
\ee
with $\gamma=2.9$ and $T_c \simeq 0.481$.  This temperature 
is indicated with a vertical line in Fig.~\ref{tau}. When 
plotted as a function of $T-T_c$ in a log-log representation,
the data for $K=24$ follow the power law divergence over
almost the entire 
temperature regime we were able to study numerically. 
We observe a power law regime over nearly 4 decades 
of slowing down, which is more than what is typically observed 
in supercooled liquids.

When $K$ is decreased, however, stronger deviations from this power law
divergence appear, and the deviations appear at 
higher temperatures when $K$ becomes smaller. 
Accordingly, the mode-coupling power law is obeyed over a range 
which shrinks with decreasing $K$, see Fig.~\ref{tau}b.
For $K=1$, the power law is followed over about 2 decades only, which 
means that the system is quite far from its mean-field limit. 
In fact, without the $K=24$ data as a guide, it would be difficult 
to argue that a mode-coupling crossover occurs at all for 
this system~\cite{xover},
a situation which we will again encounter in Sec.~\ref{md}  
when performing molecular dynamics simulations. Note that 
in this analysis, we have not tried to use $T_c$ as an 
additional free parameter, which would somewhat improve the quality 
of the power law fit.

These results clearly show that relaxation in the KFA model is a combination
of mode-coupling  and cooperative dynamics, whose respective importance depends
on the temperature, the geometry of the underlying lattice (i.e. the value 
of $K$), and the considered timescale. 
For large values of $K$ we find an apparent 
mode-coupling divergence analogous to the one reported for 
the Bethe lattice with the same connectivity. 
However, contrary to Bethe lattices, the dynamic singularity 
is eventually avoided for {\it any} value of $K < \infty$.  
This can be proven rigorously, the argument being deferred to the 
Appendix.
The proof for $K>1$ is a straightforward generalization
of the $K=1$ case discussed previously~\cite{TBF}, and it shows that 
for any finite $K$ the relaxation time 
diverges only at $T=0$, with a  divergence which 
becomes steeper when $K$ increases.
This results from the behaviour of the 
upper bound we derive for $\tau_\alpha(T)$, showing that 
$\tau_\alpha(T)$ cannot grow faster than 
$\exp[ K \exp(K/T) ] $, to leading order 
at low $T$. 

Interestingly, this argument also suggests that using any of the standard 
definitions of the kinetic fragility, the KFA model 
would become more fragile with increasing the value of $K$,
the fragility even becoming formally infinite  
when $K \to \infty$. The fragility increases with $K$ 
because the dynamics is more influenced by the mode-coupling crossover 
and closer to an algebraic singularity at $T_c$, which is responsible
for the fragility increase observed numerically in Fig.~\ref{tau}.
In the low-temperature regime below the mode-coupling crossover 
where the upper bound derived  in the Appendix is valid, 
the corresponding fragility also increases because the effective 
activation energy increases with $K$. Within the KFA model, 
we arrive to the intriguing conclusion 
that systems where the mode-coupling crossover is more
pronounced are also characterized by a larger kinetic fragility.
It would be interesting to know 
whether such a correlation holds for real supercooled liquids. 
We discuss this issue further in Sec.~\ref{md}.
 
\subsection{Dynamical finite size effects}

Having established that the KFA model smoothly interpolates between
mode-coupling and cooperative dynamics, we are now in a position to study
how the different regimes are affected by confinement and the corresponding
finite size effects. In order to do that, we shall analyze how the 
alpha-relaxation depends on the 
control parameters $(K, T, L)$. We recall that for small 
system sizes, instances containing a blocked backbone 
may appear with a finite
probability. When this probability becomes large, 
the KFA model is no longer a physical description 
of a realistic glassy liquids which cannot be truly blocked. 
In a more realistic system, 
a different relaxation mechanism will replace facilitation.  
In order to avoid this problem, we have decided to reject those instances 
and perform an average over samples which do not contain
blocked backbones. 

We find that 
the shape of the persistence functions depends weakly but in a non-trivial
manner on the control parameters. For this section we use, for 
convenience, an integral definition of the relaxation time:
\be
\tau_\alpha = \int_0^\infty dt \, P(t).
\ee
We have checked that our conclusions do not 
depend on this particular defintion of $\tau_\alpha$. 
We have explored the dependence of $\tau_\alpha$ on all three parameters
over a wide range. We now present the salient features of the 
finite size effects within the KFA model.

\begin{figure}
\psfig{file=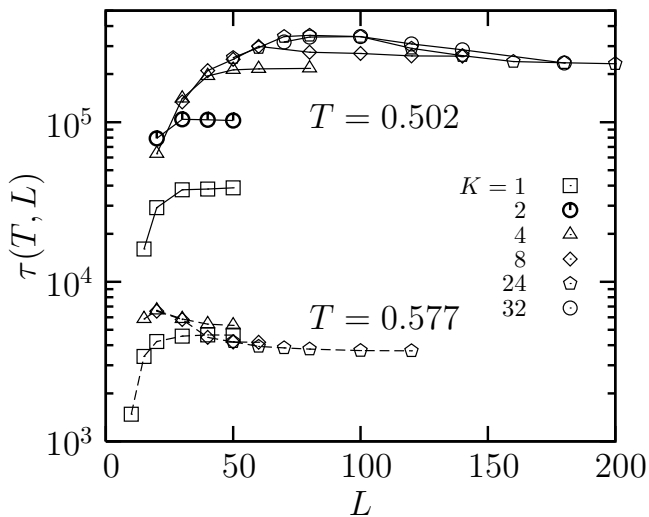,width=8.5cm}
\caption{\label{taurfar8a} System size dependence 
of the relaxation time 
in the KFA model
at two fixed temperatures and various 
values of the interaction range $K$.
The strong temperature evolution of the finite size effects 
indicates the growth of $\ell_{FS}(T)$ at low $T$. 
Note the different $L$ dependence in the cooperative 
($\tau_\alpha$ increases with $L$)
and the mode-coupling ($\tau_\alpha$ decreases with $L$) regimes.}
\end{figure}

First, we fix the temperature, and study how the relaxation time 
reaches its bulk value for $L\to\infty$ for different connectivity
ranges $K$, see Fig.~\ref{taurfar8a}. For a moderate temperature,
$T=0.577$ (recall that $T_c=0.481$) bulk relaxation times 
are of the order $5 \cdot 10^3$ for all $K$. We observe that 
this value is reached for moderate system sizes in all cases, 
$L \approx 20-40$, and that the $K \to \infty$ limit is reached very quickly 
as well, since the data do not change for $K \ge 8$. Strikingly, we find that 
the asymptotic value of $\tau_\alpha(L)$ is reached from below for $K=1$
and from above for $K \ge 4$, suggesting that 
finite size effects are qualitatively different for mode-coupling ($K>4$)
and cooperative ($K<4$) regimes at this temperature.  

These observations are amplified at lower temperature. 
For $T=0.502$, bulk dynamics is recovered only at much
larger system sizes, from $L \approx 30$ for $K=1$ to $L\approx 150$
for $K=24$. Moreover, since we are very close 
to the mean-field singularity, the $K \to \infty$ limit 
is only achieved for much larger $K$, near $K \approx 32$. 
Thus, by decreasing the temperature we observe enhanced finite 
size effects in both dynamical regimes, which constitutes direct 
evidence that dynamical slowing down is accompanied by a growing
correlation length scale, and unveils the existence of 
a growing length $\ell_{FS}(T)$, that 
determines dynamical finite size effects and 
which grows by lowering the 
temperature. We emphasize that this result holds both in 
mean-field and cooperative regimes even though the system size dependence
is qualitatively different in both cases.

\begin{figure}
\psfig{file=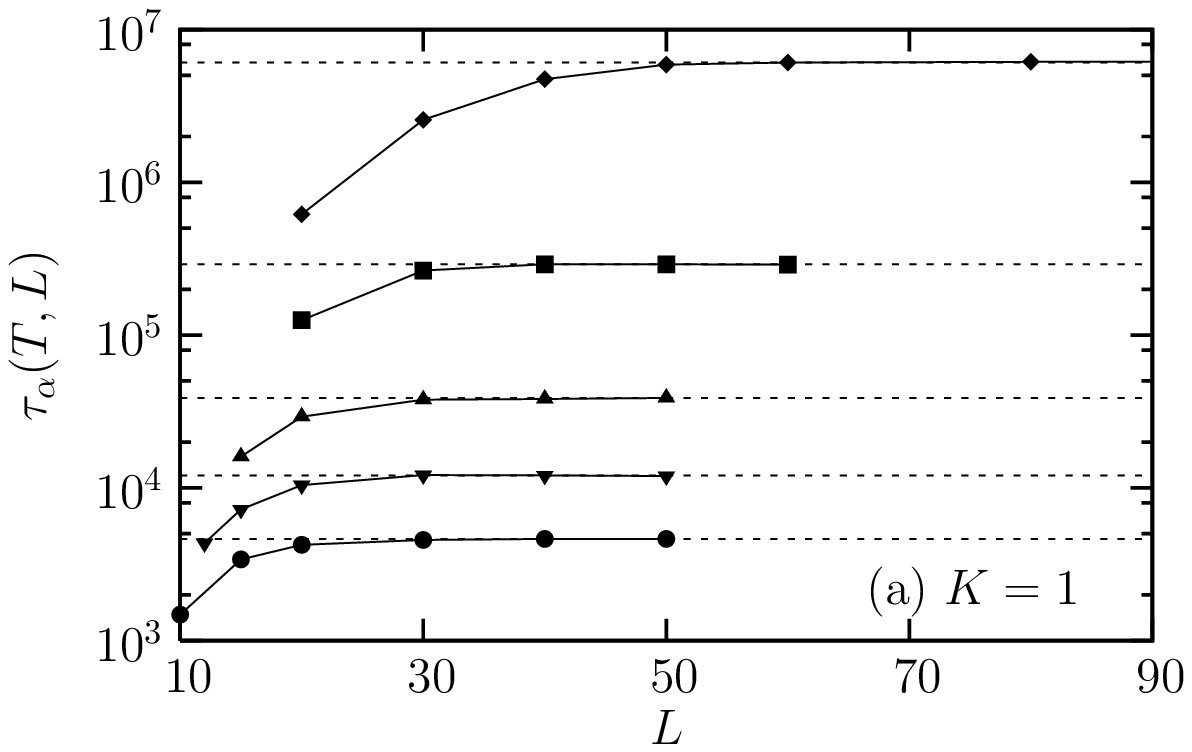,width=8.4cm}
\psfig{file=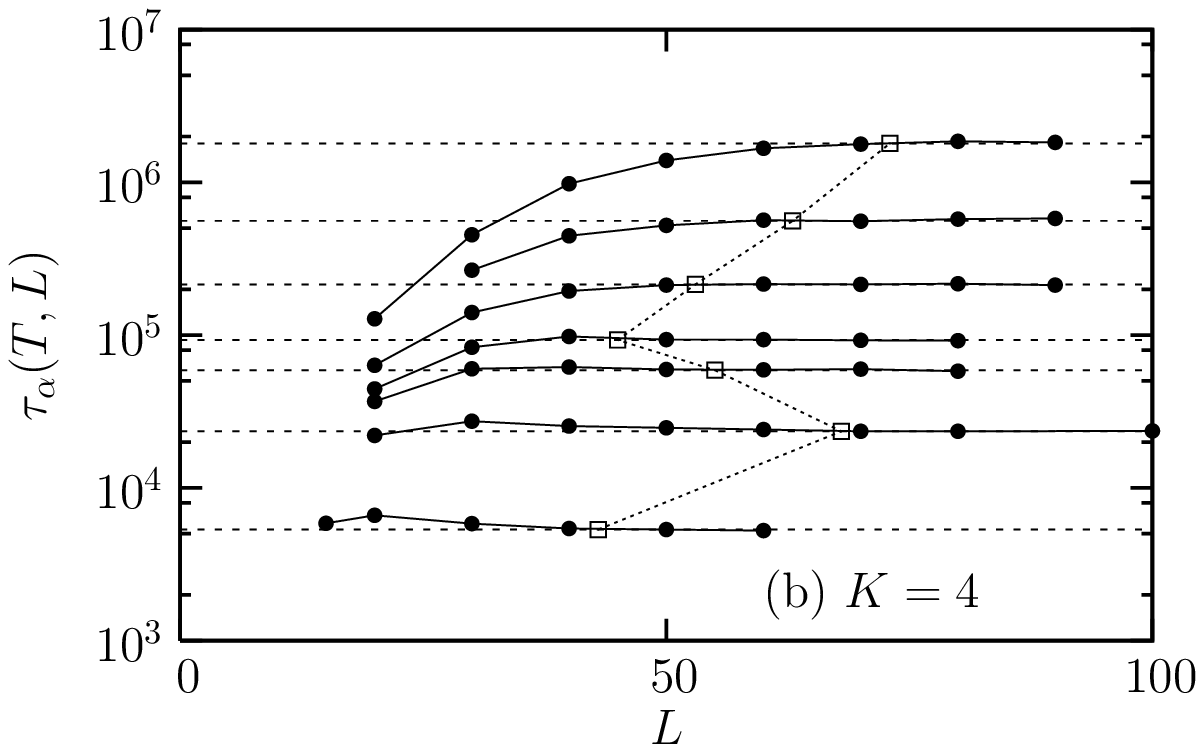,width=8.5cm}
\psfig{file=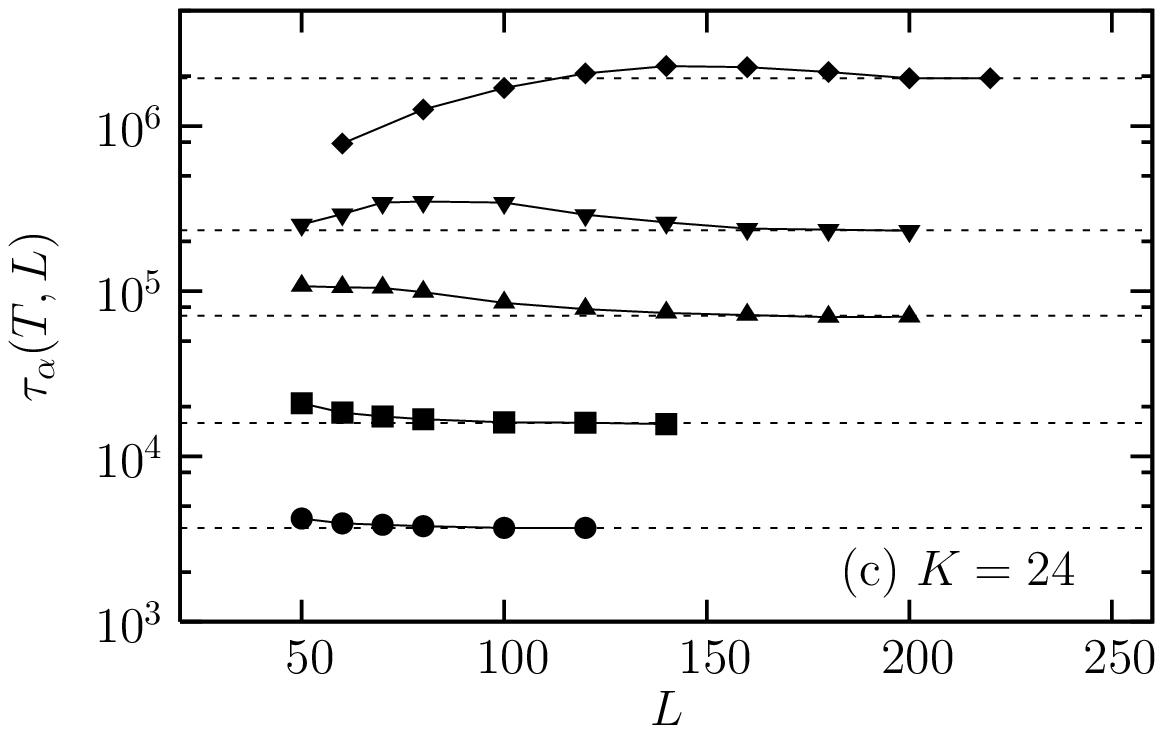,width=8.2cm}
\caption{\label{taurfar8b} 
Temperature evolution of finite size effects.
(a) $K=1$ and various temperatures, $0.577 \le T \le 0.409$, 
(b) $K=8$ and various temperatures, $0.577 \le T \le 0.490$, 
(c) $K=24$ and various temperatures, $0.577 \le T \le 0.490$.
While $\tau_\alpha$ grows with $L$ in the 
purely cooperative regime ($K=1$ at all $T$, $K=4$ at low $T$), 
it decreases with $L$ in the mode-coupling regime ($K=4$ and 24 at 
moderate $T$), and has a non-monotonic size dependence 
when both regimes coexist and compete ($K=8, 24$ near $T_c$).
This competition produces a non-monotonic 
temperature evolution of the characteristic length
$\ell_{FS}(T)$, indicated by open symbols in 
(b).}  
\end{figure} 

We now fix the connectivy range $K$, and study for each different 
$K$ value the whole temperature evolution of the finite size effects in
Fig.~\ref{taurfar8b}. 
For the case $K=1$, shown in 
Fig.~\ref{taurfar8b}a, the mode-coupling regime is almost 
absent and the dynamics appears to be `activated'
and non-mean-field in the whole slow dynamics regime. 
Correspondingly, the relaxation
time increases with $L$ for all temperatures with no sign of non-monotonic
behaviour.
This behaviour can be explained by recalling that relaxation in the $K=1$ 
model proceeds via diffusion of so-called `macro-vacancies'. 
The relaxation time corresponds to the time it takes 
a macro-vacancy to diffuse over an area comparable to the inverse of the 
macro-vacancy density. 
When the system size becomes smaller than this characteristic area 
dynamical finite size effects set in. 
The interpretation is that by sampling configurations that do not contain 
blocked structures, we are effectively
conditioning the sampled configurations to always contain at least one 
macro-vacancy. 
The area over which this macro-vacancy must diffuse 
becomes smaller when $L$ decreases, and so does 
the relaxation time. This argument also explains why convergence 
to the bulk behaviour is reached only at system sizes that 
grow with decreasing $T$, because the density of macro-vacancies
decreases. Therefore, the growth of $\ell_{FS}(T)$ is directly 
related to the growth of dynamical correlations in this regime
which directly control the finite size effects 
observed in Fig.~\ref{taurfar8b}a.

The situation for $K=4$, shown in Fig.~\ref{taurfar8b}b,
is different because mode-coupling dynamics
now controls the relaxation over an intermediate 
temperature regime, as discussed above. We have argued in Sec.~\ref{mct} that 
mode-coupling dynamics in finite systems should be slower than in the bulk, 
which is indeed compatible with the higher 
temperature data shown in Fig.~\ref{taurfar8b}b,
which show that the bulk value of the relaxation time is 
reached from above. This situation is in stark contrast with 
the cooperative behaviour obtained for $K=1$ in Fig.~\ref{taurfar8b}a.
Decreasing the temperature has two effects. 
First, bulk 
dynamics is reached only at  system sizes that grow, because 
$\ell_{FS}(T)$ grows, as noticed above. 
Second, the presence of the mode-coupling regime 
becomes evident, as dynamics becomes cooperative and thus reacts 
differently to finite sizes. At very low temperature we find 
that dynamics is fully cooperative, and $\tau_\alpha$ increases
with $L$. A striking behaviour is observed near $T \approx T_c$ 
where both types of dynamics coexist and compete to 
yield a non-monotonic behaviour of $\tau_\alpha(L)$,
which should be interpreted as a `mixture' of high and low temperature 
behaviours. 

A second striking consequence of this competition
is that the length $\ell_{FS}(T)$, which can be estimated
as the system size needed for $\tau_\alpha(L,T)$ to converge
to the bulk value, has a non-monotonic evolution with temperature, 
as indicated by the open symbols in Fig.~\ref{taurfar8b}b. 
These open symbols have been placed in between the first two consecutive $L$
values for which the relaxation time does not evolve anymore, within 
statistical accuracy. While these points do not represent the result
of the quantitative determination of a characteristic system size, 
they describe qualitatively well the 
numerical data. This behaviour occurs near the mode-coupling temperature 
and the minimum of $\ell_{FS}(T)$ occurs when 
the opposite effects of cooperative and mode-coupling dynamics
on the relaxation time nearly compensate to produce 
negligible finite size effects. The effective non-monotonic 
temperature evolution of $\ell_{FS}(T)$ is strongly 
reminiscent of the numerical findings of Ref.~\cite{sandalo},
as we discuss further in Sec.~\ref{conclusion}.
By contrast, the four-point dynamic susceptibility 
measured in the bulk does not show such a non-monotonic 
temperature evolution, as discussed in the Appendix.

We were only able to detect such a striking non-monotonic temperature 
evolution of $\ell_{FS}(T)$ over a narrow range, $4 \le K \le 8$.
This is because small $K$ values are little influenced
by the mean-field limit, while for large $K$ we cannot 
study low enough temperatures and enter the fully cooperative 
regime. Indeed, for $K=24$ shown in Fig.~\ref{taurfar8b}c, 
we find a qualitatively similar 
coexistence and non-monotonic behaviour at intermediate 
temperatures, and the effect is even more pronounced 
because for this $K$ value, the mode-coupling singularity is 
only narrowly avoided, and the mode-coupling regime 
extends to much lower temperatures. However, for this value
we have not been able to reach sufficiently low temperatures to see 
purely cooperative dynamics and a monotonic increase of $\tau_\alpha$
with $L$, see the lower temperature in Fig.~\ref{taurfar8b}c.

In conclusion, the study of finite size effects 
in the KFA model, where the relative importance of mode-coupling
and cooperative dynamics can be controlled, 
supports the validity of the theoretical arguments developed
in Sec.~\ref{theory2}. We find in particular that 
both temperature regimes exhibit qualitatively distinct 
response to the use of finite sizes, while in the crossover 
region a remarkable non-monotonic size dependence 
is obtained, which reveals in a very direct manner that
the nature of the relaxation is changing near the avoided 
singularity $T_c$. We also observe that for systems 
that are too far from the mean-field limit, this crossover 
is too weak, dynamics is mostly controlled by 
cooperative processes and $\tau_\alpha$ increases monotonically with $L$. 
Therefore, finite size effects can be viewed 
as a powerful tool to probe the existence of a physically 
relevant temperature regime where mean-field-like dynamics 
prevails.   
In all cases, we find that finite size effects appear for $L \le \ell_{FS}(T)$,
where $\ell_{FS}(T)$ represents a length scale which grows upon 
decreasing the temperature in both mode-coupling and cooperative regime, 
but exhibits an apparent non-monotonic temperature evolution 
when the opposite effects of mean-field and cooperative 
dynamics nearly compensate.

\section{Results from molecular dynamics simulations}
\label{md}

\subsection{Models and bulk behaviour}

Given the diversity of behaviours predicted 
theoretically in previous sections, we have decided to 
undertake a large numerical effort to 
investigate finite size effects in a broad variety of 
model systems using molecular dynamics simulations and 
to study liquids with different interactions and 
kinetic fragilities. 
We have performed large-scale 
simulations of four model liquids
representative of different classes of systems.

The first model, which we call the `network liquid', was introduced
and studied in Ref.~\cite{coslo}. Using carefully chosen 
Lennard-Jones interactions between the two components 
of a binary mixture, it is possible to mimic
the structure of network-forming liquids (such as silica)
while avoiding the use of long-range electrostatic 
interactions, which is especially convenient when small systems
need to be studied. 
Additionally, at low enough 
temperatures the temperature dependence of the relaxation time 
was found to 
be close to an Arrhenius law~\cite{coslo}, and 
therefore we use this
network liquid as representive of the class of strong 
glass-formers. 

The second model we study is the binary Lennard-Jones mixture 
introduced in Refs.~\cite{KALJ}. The model was originally 
devised as a simple Lennard-Jones model for
two-component metallic glasses and has become a canonical 
system for numerical studies of supercooled liquids~\cite{andersen}. 
Its relaxation time 
grows in a super-Arrhenius manner, and so it is considered 
as a good model for fragile liquids. Although comparing kinetic fragilities 
between simulations and experiments is not straightforward, 
the binary Lennard-Jones 
mixture has an `intermediate' fragility, which suggests it is less 
fragile than typical fragile glass-forming materials 
studied in experiments such as for instance 
ortho-terphenyl~\cite{coslo,gilles}. 

The third model is also a canonical model for studies 
of the glass transition. It is a binary mixture of soft spheres
interacting with a purely repulsive $r^{-12}$ potential
introduced in Ref.~\cite{bernu}. Its behaviour is 
in fact very similar to the one of the binary Lennard-Jones
potential, since this model also seems to display an intermediate 
kinetic fragility. 

The fourth model we study is a binary mixture of soft 
repulsive particles interacting with a one-sided 
repulsive harmonic potentiel. The model was introduced 
in Ref.~\cite{durian} as a model for wet foams, and its
glass-forming properties were studied in Refs.~\cite{tom},
where it was shown that over a broad regime of densities, 
this system actually behaves as a quasi-hard sphere system.
Comparing the kinetic fragility of hard spheres (whose glass transition
is controlled by density) to molecular liquids (controlled 
by temperature) is ambiguous~\cite{tom}. However, using 
the compressibility factor $Z = P/(\rho T)$ to build the analog
of an Arrhenius plot for hard spheres~\cite{tom} suggests that 
hard spheres are actually characterized by a rather large kinetic 
fragility. We  take harmonic spheres as being representative 
of fragile glass-forming materials.  

Because these models have been studied extensively before, we
only provide limited details about our simulations 
in the sections below, and refer
to the original publications for more informations. 
Our focus in this work was to analyze how simulations 
in finite size systems differ from the bulk behaviour, 
and whether the observed finite size effects could be interpreted
along the lines discussed in previous sections.  
As a dynamical observable we measure the behaviour of the self-intermediate
scattering function, $F_s(q,t)$, 
and determine the alpha-relaxation time $\tau_\alpha(T)$, from the time-decay 
of this correlator to the value $1/e$ and a wavevector 
corresponding to the first peak of static structure, as is 
usual~\cite{walterhouches}. 
Our central aim is to measure 
$\tau_\alpha(T,N)$ for systems containing a finite number of particles,
$N$. 

\begin{figure}
\psfig{file=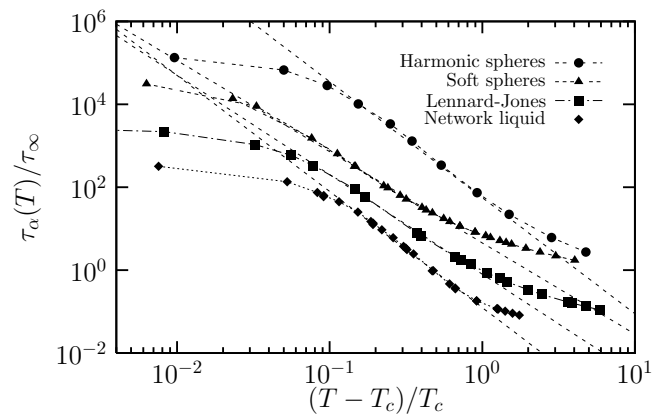,width=8.5cm}
\caption{\label{mctfits} Bulk relaxation time $\tau_\alpha(T)$ 
of the four model liquids studied in this work using the 
representation of Fig.~\ref{tau}b appropriate for detecting 
a mode-coupling algebraic divergence at temperature $T_c$. 
We vertically shift the systems by a time constant $\tau_\infty$
for clarity, and show as a dashed line the result of 
power law fits with exponents $\gamma = 2.8$, 2.4, 2.2  and 2.8
for the network liquid, Lennard-Jones particles, soft and harmonic 
spheres, respectively. The power law fit is obeyed over a broader range from 
bottom to top.}
\end{figure}

By construction, simulations are performed in a temperature regime
which corresponds to the first 4-5 decades of dynamical
slowing down. Thus, this regime is typically the one
where predictions from the mode-coupling theory are usually 
tested. Therefore, our simulations fall in the temperature 
range where a crossover from mode-coupling to activated dynamics
might occur. In Fig.~\ref{mctfits} we give evidence that 
such a crossover seems to be present in the bulk dynamics
of the four liquids. For all liquids, we measure the 
temperature dependence of the bulk relaxation time. We then fit
its temperature evolution to a power law divergence, as in 
Eq.~(\ref{powerlaw}), to estimate 
the location $T_c$ of the mode-coupling 
singularity. We present the data for the four liquids in 
Fig.~\ref{mctfits} using the same representation as in Fig.~\ref{tau}b,
where a power law divergence appears as a straight line. 
For all liquids, a power law regime is obtained for intermediate 
temperatures, although the time window over which it applies
depends on the particular system. Unsurprisingly we find 
that a power law divergence is not very pronounced for the 
network liquid which rapidly enters an Arrhenius regime at low 
temperatures, while the harmonic sphere system is the one 
for which the power law is the most convincing. The 
Lennard-Jones and soft sphere mixtures have an intermediate behaviour.
Thus, we find that the degree to which mode-coupling theory predictions 
apply (at least for the bulk relaxation time) seems 
correlated with the kinetic fragility of the model. The same connection
was found in the KFA model in Sec.~\ref{rfa}.  
We emphasize that a power law fit to the relaxation 
for moderately supercooled liquids is bound to yield 
a quantitative estimate of the value of $T_c$, but this 
does not necessarily imply, as we shall see, that the long-time dynamics
is truly controlled by the mode-coupling physics~\cite{xover}.

\subsection{Network liquid with Arrhenius behaviour}
\label{network}

We start our discussion with the results obtained for the network
liquid~\cite{coslo}. 
The model is an ${\rm AB}_2$ binary mixture designed to be a simple 
analog of silica, ${\rm S_i O}_2$,  
forming a connected assembly of tetrahedric structures. 
For this system, we find that dynamics becomes slow when 
temperature becomes smaller than 
$T \approx 0.45$, and we can follow finite size effects down to 
$T = 0.29$, where bulk dynamics has slowed down by about 4 decades.
Using a power law fit of these data, we obtain 
an exponent $\gamma \approx 2.8$ and extract the location of the 
mode-coupling temperature, $T_c \approx 0.31$. Since the power law
is obeyed over a limited temperature range, it is 
relatively easy to access
temperatures lower than $T_c$ in equilibrium conditions, 
as found also in a more realistic model of silica~\cite{horbach}.
The density of the system, $\rho=1.655$, has been adjusted 
to best reproduce the structure of silica obtained 
in molecular dynamics simulations performed at density 
$\rho = 2.37 {\rm g}/{\rm \AA}^{3}$~\cite{horbach}.

For this system we performed simulations both using a thermostat 
(in the NVT ensemble~\cite{nose}), 
or without thermostat after proper thermalisation
(in the NVE ensemble), because as 
discussed in Refs.~\cite{jcp} thermally activated processes
for systems evolving with Newtonian (i.e. energy conserving) 
dynamics might induce dynamical correlations between particles
when the heat needed to cross a barrier locally is borrowed 
to neighboring particles. We found no quantitative differences 
between the two sets of simulations, and we have therefore 
merged the two sets of simulations for 
the bulk data presented in
Fig.~\ref{mctfits}.

\begin{figure}
\psfig{file=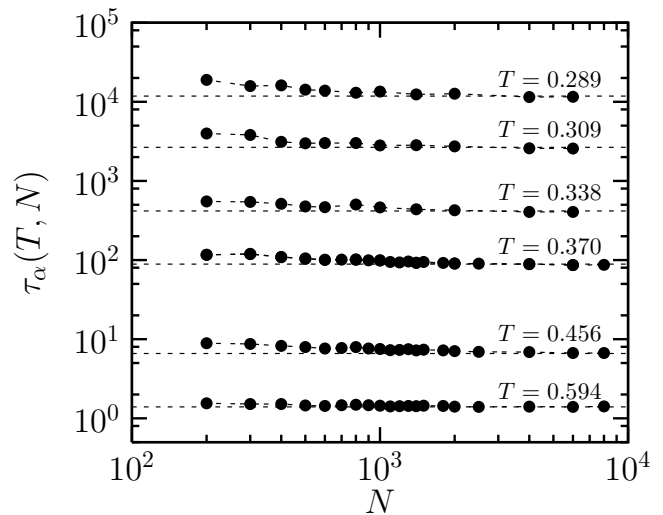,width=8.5cm,clip}
\caption{\label{daniele} System size dependence of relaxation time in 
the network liquid where $T_c \approx 0.31$. 
The dynamics slows down for smaller systems, but the amplitude
and range of the effect evolves weakly with temperature.}
\end{figure}

The results corresponding to dynamical finite size effects are reported 
in Fig.~\ref{daniele}, where both 
NVT and NVE simulations are shown, yielding quite similar results. 
We use system sizes that are limited on the small $N$ side by the fact 
that the static structure, as revealed by the pair correlation 
function $g(r)$, becomes sensitive to $N$ and the network of tetrahedra
does not fit well the small 
simulation box. For large $N$, we easily observe convergence
to the bulk behaviour for the relaxation time as soon as $N$ is 
larger than a few thousands particles.   
In Fig.~\ref{daniele} we observe a small finite size 
effect, since $\tau_\alpha(T, N)$ reaches its bulk value
from above, that is, small systems are slower than larger ones. 
However, this effect is quite modest
and, more importantly, it does not seem to evolve very much over
the temperature range where slow, Arrhenius 
dynamics is observed, i.e. $T \le 0.45$.
Therefore, for the network liquid, we find no clear 
evidence that the length scale $\ell_{FS}(T)$ becomes large 
at low temperatures. 
These results are consistent with the view that, 
for systems showing an Arrhenius behaviour,
relaxation remains `local' and 
that correlations are rather weak and evolve very slowly 
with the temperature~\cite{heuerbks}. 

Dynamic heterogeneity has not been 
studied in detail for the present network liquid, but 
growing four-point dynamic susceptibilities  (but not dynamic
length scales)
have been reported in numerical studies of silica~\cite{vogel,berthiersio2}.
A way to reconcile these findings with our result is either that 
$\ell_{FS}(T)$ is not related to the dynamical correlation length, or
that dynamic susceptibilities grow at low temperatures because the 
strength, rather than the spatial extent, of dynamic correlations
increase at low $T$, a point that deserves further studies.  

Contrary to what was found for the activated (and highly cooperative)
regime of the KFA model,
the small dynamical finite size effect 
reported for the network liquid goes in the opposite direction of making 
the dynamics slower for smaller systems, which indicates 
that this finite size effect is not related to a competition between 
the system size and the spatial extent of the relaxaxing `entities'.
A possible explanation is that for small system sizes, the silica-like 
network is somewhat frustrated by the periodic boundary conditions, 
which could increase slightly the energy 
barrier needed to form a defected tetrahedra, and thus slow down 
the dynamics. In this view, 
dynamical finite size effects would be dominated 
by non-collective effects and, possibly, related to a 
subtle change of the static structure, an issue worth pursuing. 

\subsection{Binary Lennard-Jones mixture}
\label{lj}

We now turn to the case of the binary Lennard-Jones mixture~\cite{KALJ}. 
This model displays super-Arrhenius relaxation, and 
it is a model for which several quantitative tests of the scaling predictions 
of the mode-coupling theories have been performed with some 
success~\cite{KALJ}, even though deviations from the predictions 
can be observed at low temperatures~\cite{jcp}. 
Additionally, growing dynamic length scales have been reported for
this system~\cite{donati,donati2,berthier04,karmakar,dalle_07}. For all 
these reasons, one may expect
a more interesting temperature dependence of finite size effects
for this model. 
 
\begin{figure}
\psfig{file=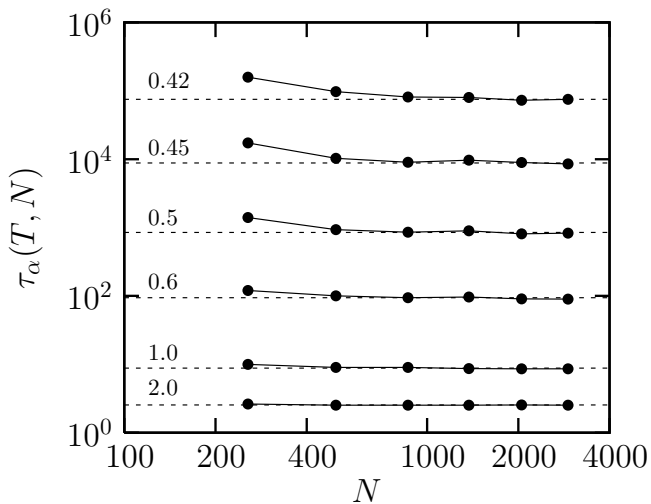,width=8.5cm}
\caption{\label{LJ} System size dependence of relaxation time in 
a Lennard-Jones liquid from the high temperature liquid, $T \ge 1.0$
down to below the mode-coupling temperature $T = 0.42 < T_c \approx 
0.435$.} 
\end{figure}

The results of our molecular dynamics simulations 
are shown in Fig.~\ref{LJ}. The simulations were performed in the 
microcanonical ensemble only, the value of the total energy
being carefully controlled in each independent 
simulation to maintain the temperature equal to the desired value,
and prevent spurious fluctuations in the dynamics. 
We present data for the high temperature liquid, $T=2.0$, 
and below the onset of glassy dynamics, $T \approx 1.0$, 
down to $T=0.42$ which lies below the fitted value of the mode-coupling 
temperature $T_c \approx 0.435$~\cite{KALJ}. 

We find that almost no dynamical finite size effects are present
above and near the onset temperature for slow dynamics,
consistent with the idea that, in simple liquids, 
relaxation is a fast local process.
For $0.5 \le T \le 1.0$, we find that finite size effects 
are present at small sizes and that the dynamics slows down
when $N$ is small~\cite{heuer1}. Remarkably, we find that this effect 
becomes more pronounced both in amplitude and in range, 
suggesting that the interplay between system size and 
structural relaxation has a more collective nature than 
in the Arrhenius liquid studied in Sec.~\ref{network}.
This suggests that a non-trivial characteristic length scale
$\ell_{FS}(T)$ grows when temperature is decreased below 
the onset in this model, in agreement with previous 
work~\cite{dasguptafss,proc}. Repeating the empirical 
analysis performed recently in Ref.~\cite{proc}, we find similarly 
that the typical lengthscale over which finite size
effects occur grows by about 50\% over the 
temperature range $T=0.42 - 1.0$. 
Contrary to previous work~\cite{dasguptafss}, however, we 
always find a monotonic $N$ dependence of $\tau_\alpha$, even at 
low temperatures. We believe that the (relatively small) non-monotonic
size dependence reported earlier~\cite{dasguptafss} was due  
to statistical uncertainty~\cite{karmakar2}.

%Additionally, when performing simulations below the  
%mode-coupling temperature, $T=0.42 < T_c \approx 0.435$,
%we find that $\ell_{FS}(T)$ does not increase much further, 
%which could indicate that its growth becomes less rapid at very low
%temperatures. This behavior is reminiscent of the one of
%four-point dynamic susceptibilities which also display a
%change in the temperature dependence across the mode-coupling 
%crossover~\cite{jcp,dalle_07}. 

Another result obtained below $T_c$ is that we 
do not find a qualitative change in the size dependence
of the relaxation time and the bulk value is still reached
from above  by increasing $L$ for 
this low temperature. Therefore, 
contrary to what has been found for the KFA model 
or predicted on general grounds for activated relaxation, 
we do not find any non-monotonic behavior near or below $T_c$.
We find this result somewhat surprising and suggest several 
hypothesis to account for these observations. First, it could be 
that the mode-coupling crossover is absent or very weak 
in this case, as in the network liquid studied above. This 
is however at odds with previous work establishing 
the validity of the scaling predictions of mode-coupling theory for this
system at intermediate temperatures~\cite{KALJ}. 
The second hypothesis 
is that activated dynamics at low temperatures involves 
a cooperativity length that has not yet grown very large, 
and thus cannot compete with the system size $L$ before other,
more microscopic effects, also appear such that there is no
room for the general argument of Sec.~\ref{theory1} to apply. 
By this argument we would conclude that much lower temperatures 
should be studied to reveal a change in the nature of finite size
effects in this model, which is at present beyond our numerical 
capabilities. A possible interpretation is that 
static point-to-set 
correlation length scales do not grow significantly in this 
system over the temperature regime currently accessible to simulations, 
or at least much less than dynamical correlation length scales.
See Refs.~\cite{glen,coslovich_07,coslovich_11} 
for recent work on static correlations in this system.

\subsection{Soft spheres}
\label{ss}

In this section we study the binary system of soft spheres 
introduced long ago in studies of the glass transition~\cite{bernu}.
We choose the particular  
model studied by several groups~\cite{bernu,roux,cavagna,yamamoto}, 
namely a 50:50 mixture of 
soft spheres interacting with an $r^{-12}$ repulsion, 
different species having different sizes. We use 
a diameter ratio 1.2 and adopt the same units as in 
Ref.~\cite{cavagna}, where density $\rho$ is fixed and temperature $T$ 
decreased (although this is a matter of convenience for this system
since the static structure is 
uniquely controlled by the combination $\Gamma=\rho T^{-1/4}$).
It was shown that this model displays in the supercooled regime 
increasing dynamic~\cite{yamamoto} and static~\cite{biroli_08} length scales. 
We perform simulations in the canonical NVT
ensemble, using a Nos\'e-Poincar\'e thermostat
with inertia parameter $Q=5.0$~\cite{nose}.

A new technical difficulty for this system is that its glass-forming 
ability is worse than the three other models studied in this work. 
In particular, we found that crystallization intervenes
very easily when temperature is decreased, especially in 
small systems. Thus, we had to carefully determine for each independent
sample whether it had crystallized in the course 
of the simulation or not. We did so by measuring several structural 
indicators, such as pressure, energy, and pair correlation function
from which crystallization was obvious. 
Therefore, in the data presented below, we only 
consider state points where crystallization 
was found very infrequently. In practice, we do not show 
data when crystallization occurred in more than $30~\%$ of the samples. 

A few of the remaining 
data are still a little ambiguous, as we observe 
fluctuations of the potential energy that are large and long-lived,
but do not correspond to an irreversible crystallization 
of the system. This is reminiscent of the numerical
observations reported for another binary mixture~\cite{pedersen}.
For these runs, dynamics is typically slower than the average, 
and it is not clear whether these runs should be discarded 
(as being affected by incomplete crystallization), or 
averaged together with `normal' samples
(as being characterized by some other forms of 
local ordering).
We checked that the main conclusions reported below
are not affected if we remove these very slow samples.

\begin{figure}
\psfig{file=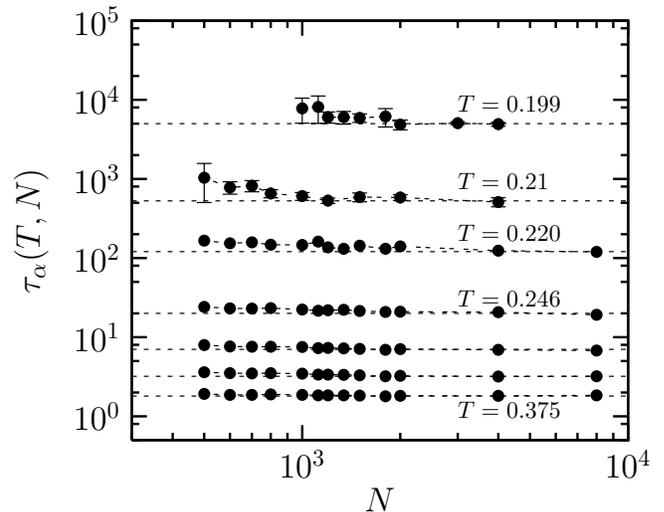,width=8.5cm,clip}
\caption{\label{daniele2} 
System size dependence of relaxation time in 
a soft sphere mixture from the high temperature liquid, $T \ge 0.25$
down to the mode-coupling temperature ($T_c \approx 
0.198$). We only consider state points where crystallization is 
very infrequent. The overall temperature evolution 
is similar to the Lennard-Jones results in 
Fig.~\ref{LJ}.} 
\end{figure}

For this system, the bulk data were fitted to a power law divergence, 
and the result of this fitting is shown in Fig.~\ref{mctfits},
where we use $\gamma \approx 2.2$ and $T_c \approx 0.198$.
We note that this value is significantly smaller than 
the values ($T_c = 0.226 \ldots 0.246$) quoted in the 
literature~\cite{roux,cavagna,cavagna2,cavagna3}, 
which stems from very early work~\cite{roux}, and presumably 
overestimated the mode-coupling temperature 
by a very large amount.

We present our results for finite size effects in the soft sphere mixture 
in Fig.~\ref{daniele2}. As for the Lennard-Jones model, 
we find that dynamics is rather insensitive to system sizes in the high
temperature liquid, but becomes size dependent below the onset 
of glassy dynamics, which we locate near $T \approx 0.25$. 
The size dependence also extends to larger sizes when $T$ decreases,
signalling again the growth of the characteristic length 
$\ell_{FS}(T)$ with decreasing $T$. Unfortunately, 
due to the crystallization issue mentioned above, it is not easy to 
follow the size dependence to very low temperatures over a broad 
range of system sizes. The limited amount of data shown 
in Fig.~\ref{daniele2} seems to suggest that soft spheres 
have a behaviour similar to the one observed in the Lennard-Jones 
system. In particular, the size dependence for $T=0.2$,  
near the mode-coupling temperature, does not show sign 
of a qualitatively different behaviour as compared to
higher temperatures. We were not able to  study this system at even 
lower temperatures, because of crystallization issues. 
We suspect in particular that the very strong finite size effects 
reported in Ref.~\cite{kim} might be affected by crystallization as 
well, since the size dependence reported in Fig.~\ref{daniele2} 
is more modest. 
 
\subsection{Harmonic spheres}
\label{harmonic}

The final model we consider is a 50:50 binary mixture of 
harmonic spheres with diameter ratio 1.4, which we study using 
molecular dynamics simulations in the microcanonical ensemble. 
We use the same parameters as in Ref.~\cite{sandalo}, and work 
at constant density $\rho=0.675$ and use temperature as a control 
parameter. For this density, the onset of glassy dynamics is near
$T \approx 13$ and the mode-coupling temperature 
used in Fig.~\ref{mctfits} is $T_c \approx 5.2$~\cite{sandalo,tom}.
In contrast with the soft sphere model studied in the previous section
we find that crystallization is not an issue for systems as small as
$N=108$ over the entire range of temperatures. 
Since the range of the potential is equal to the particle 
diameter (as for hard spheres), we can in principle study
even smaller system sizes. We have found that this is only possible
for large enough temperatures, systems with $N=32$ becoming
very heterogeneous at low temperatures. 
Therefore, we shall only display data for those state points where
stability is never an issue.

\begin{figure}
\psfig{file=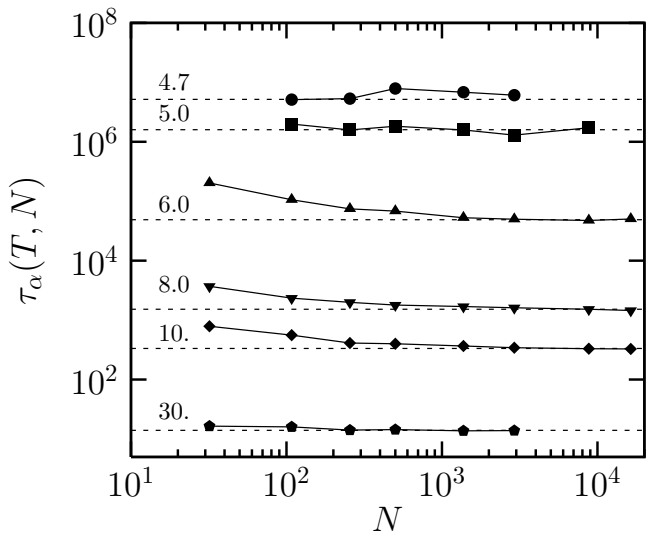,width=8.5cm}
\caption{\label{hs} 
System size dependence of relaxation time in 
a harmonic sphere mixture from the high temperature liquid, $T \ge 13$
down to the mode-coupling temperature $T_c \approx 5.2$.
Note the qualitative change of size dependence near the mode-coupling crossover
and the non-monotonic size dependence at low $T$.}
\end{figure}

In Fig.~\ref{hs} we present our results for the finite size 
dependence of the relaxation time in harmonic spheres 
across the mode-coupling crossover. For high and moderately
low temperatures, we find results that
are qualitatively very similar to the ones discussed in the previous 
sections for soft and Lennard-Jones particles, 
with no size dependence above the onset, 
a slowing down at small sizes between the onset and mode-coupling 
temperatures which becomes more pronounced if $T$ is lowered.

Strikingly, we find that near and below the mode-coupling temperature
the size dependence changes its qualitative form to become 
nearly size independent at $T = 5.0$, and even 
non-monotonic with $N$ at the lowest temperature we have been able to study,
$T=4.7$. For this temperature, we note that the maximum 
of $\tau_\alpha(N)$ occurs for a system size of about $N 
\approx 600$, where the structure is very stable
and very close to that found in the bulk system. 
Additionally, these data have been averaged 
over a large number of realizations and very long simulation 
times to reduce the statistical noise, and so the effect
reported in Fig.~\ref{hs} is physical.

The data presented in Fig.~\ref{hs} are qualitatively similar to the one
obtained for the KFA model
at intermediate $K$ (see Fig.~\ref{taurfar8b}b) 
and presumably have a similar  physical origin.
A natural interpretation of this non-monotonic 
behaviour comes from the fact that
it occurs very close to the fitted mode-coupling temperature, $T_c\approx
5.2$, where deviations from mode-coupling predictions 
are already present,
see Fig.~\ref{mctfits}. Therefore, we attribute this change 
of behaviour to a change of physical mechanism controlling the relaxation 
from mode-coupling to activated dynamics.

From the behavior shown in Fig.~\ref{hs} we conclude also that 
mode-coupling and activated 
dynamics interact and compete to produce an 
apparently non-monotonic temperature evolution of 
$\ell_{FS}(T)$, having a maximum near $T\approx 6.$ and a minimum
near $T\approx 5$, see Fig.~\ref{hs}. It is remarkable that 
this qualitative evolution with temperature 
of $\ell_{FS}(T)$ follows very closely the behaviour 
reported for dynamic profiles near an amorphous wall in 
Ref.~\cite{sandalo} for the same system. Thus, we think that the bulk data
reported in Fig.~\ref{hs} provide both an independent
confirmation and a natural physical interpretation 
of the surprising non-monotonic dependence of 
dynamic length scales found near $T_c$ in this system~\cite{sandalo}.

Finally, for the system of harmonic spheres, we have additionally 
studied the dynamics using Monte-Carlo simulations, using 
the same implementation as in Ref.~\cite{montecarlo}.
For the present system we found that Monte-Carlo
dynamics is slightly less efficient than  molecular dynamics, so that 
getting data comparable to those shown in Fig.~\ref{hs}
is challenging. Instead, we performed 
very long simulations for only a few selected
state points $(N,T)$, which confirmed that also for Monte Carlo
dynamics, the size dependence of the relaxation time
becomes non-monotonic at low temperature, $T=4.7$, and nearly size
independent at $T=5$, in agreement with Fig.~\ref{hs}. Therefore,  
this effect is not due to the specific type of dynamics 
chosen to perform our study. This also confirms that 
hydrodynamic effects~\cite{bocquet,jackle} 
play very little role in the results presented in this work.
 
\section{Summary and conclusion}

\label{conclusion}

To conclude this article, we wish to summarize the main 
results obtained in this work. First, we discussed 
from a theoretical point of view 
the possible effects and the interest of using small sizes 
to study the dynamics of supercooled liquids. We presented 
the following arguments, which depend on the precise mechanism 
envisioned for structural relaxation in systems 
approaching the glass transition.

\begin{enumerate}

\item Only minor  finite size effects are expected for 
strong glass-formers whose relaxation time 
follows an Arrhenius law because the corresponding
activation energy likely corresponds to a `local' excitation. 
Thus, the length $\ell_{FS}$ should not grow with decreasing temperature
and the relaxation timescale for small system sizes should be dominated  
by non-universal effects affecting the local energy barrier for relaxation. 

\item For cooperative, thermally activated processes, 
dynamics becomes faster if the system size decreases
because cooperative events 
then involve a smaller number of particles, thus reducing
the barrier for relaxation. In this case the growth of $\ell_{FS}$ 
should track the one of the length scale measuring cooperativity 
(e.g., the point-to-set length within RFOT).

\item Mode-coupling relaxation becomes slower in smaller systems
because spatially extended, unstable relaxation modes become 
stable in small systems \cite{silvio07}. This trend holds  
until activated dynamics takes over 
when all unstable modes have disappeared,   
and presumably makes relaxation faster as described in item 2. 
Overall, the size dependence can thus be non-monotonic
at intermediate temperatures. Moreover, this can lead to a quite unusual 
behavior of $\ell_{FS}(T)$ that would track the one of the 
dynamical correlation length.  

\item For diffusing point defects, dynamics becomes slower when 
system size decreases because another relaxation channel
must be used when no defects are present in small systems.
If cooperative activation occurs, then the dynamics may accelerate 
at small sizes, making the overall size dependence non-monotonic.  
In this case $\ell_{FS}(T)$ is expected to be related in a power law way to 
the dynamical correlation length.  

\item For kinetically constrained models, defects are the 
only channel available. Thus, dynamics becomes non-ergodic
in samples containing no defects, which can be seen as an extreme 
slowing down. By discarding these instances, one biases 
the statistical weight towards configurations with larger
concentration of defects, and as a result the measured relaxation time 
decreases when the system size decreases (the effective defect concentration
increases). 

\end{enumerate}

We have also introduced 
a new lattice glass model, the Kac-Fredrickson-Andersen 
(KFA) model, for which the distance to the mean-field (or mode-coupling like)
limit can be controlled by tuning the range $K$ 
of the spin connectivity.
We have provided numerical and analytical evidence
that this approach successfully generates an avoided mode-coupling 
singularity, in analogy with real supercooled liquids. 
The detailed analysis of the dynamical finite size effects of this model
agrees with the general theoretical predictions. For $K=1$, 
we obtain a monotonic growth of the relaxation time with system size, 
explained by mechanism 5.
However, for intermediate values of $K$, the system 
exhibits an MCT crossover and  
the behaviour follows a non-monotonic size dependence.
Although here 
the reason for non-monotonicity is not that small systems have 
activated dynamics, the study of the KFA model is a concrete example for which 
one finds that the interplay between two competing relaxation mechanisms  
can lead, for intermediate $K$ values, to a surprising non-monotonic
temperature evolution of the characteristic length $\ell_{FS}(T)$.

Subsequently, we have presented the results of 
simulations of 
four models for supercooled liquids.
Mechanism 1 gives a good description of the size 
dependence for the strong, 
network-forming liquid. The effect 
of the mode-coupling crossover described by mechanism 3 is observed in 
a model quasi-hard spheres, while the behaviour of the 
Lennard-Jones and soft spheres 
models appeared somewhat intermediate between mechanisms 
1 and 3, and was harder to interpret.  

Overall, the simulation of fragile systems seem to confirm 
the RFOT result of Ref.~\cite{silvio07}, recently discovered in 
simulation of model systems \cite{sandalo}, that dynamic and static 
length scales are largely decoupled in the mode-coupling regime
and have distinct temperature dependences, with static 
point-to-set length scales starting to show a significant 
growth at temperatures near the mode-coupling crossover, while 
dynamic length scales grow rapidly even 
at higher temperatures. A natural interpretation 
of the non-monotonic size dependence found in 
Fig.~\ref{hs} is that both types of mechanisms compete near the mode-coupling
temperature. This competition has also been 
invoked to interpret the non-monotonic
behaviour of dynamic profiles near an amorphous wall in 
Ref.~\cite{sandalo}. We have shown that a similar non-monotonic
temperature evolution of $\ell_{FS}(T)$ is obtained in the
same harmonic sphere model. 

These results contribute to a clarification 
of the nature of the mode-coupling crossover, 
and show that the strength of the mode-coupling 
relaxation mechanims depends on the specific model, and is very weak
in glass-formers with low and intermediate fragility. It would
be desirable to understand better why the mode-coupling crossover 
is more pronounced in harmonic (and presumably hard) 
spheres than in other models, in order 
to observe similar qualitative changes in the 
mechanisms responsible for structural 
relaxation for other systems and in experimental work.
Another issue worth exploring is the idea that 
using finite sizes may perturb the static structure 
of the liquid at the level of high-order correlation 
functions, which in turn could affect the dynamics.  

\acknowledgments
We thank J.-P. Bouchaud for discussions.
Some of the simulations have been carried out in HPC@LR.
C. T.  acknowledges funding from ERC Advanced Grant PTRELSS 228032.
G. B.  acknowledges funding from ERC Grant NPRGGLASS.
W. K. acknowledges support from the
Institut Universitaire de France.

\appendix

\section{More results on the KFA model}

\subsection{Relaxation time}

In this section we discuss the relaxation time
of the KFA model, showing the absence of a 
finite temperature dynamic singularity 
for $K < \infty$.

The relaxation mechanism for the square lattice $K=1$ has been discussed 
in Ref.~\cite{TBF}. 
It is explained in terms of the diffusion of macro-vacancies, i.e.
extended defects.
The relaxation timescale is given by 
the time it takes to macro-vacancies to diffuse over an area proportional to 
the inverse of their density. 
Since the diffusion coefficient of a macro-vacancy
simply leads to subleading corrections, one finds that the relaxation 
timescale is given by the inverse
of the probability of having one macro-vacancy around a given site. 
This reads \cite{TBF}:
\be
P = \prod_{\ell=1}^\infty 
[ 1 - (1-c)^\ell]^4 \sim e^{-4/c} \sim e^{-4 \exp(1/T)} > 0.
\ee
This argument shows that the relaxation timescale cannot be larger than 
$1/P$, which is a finite number at all $T$. Therefore, the $K=1$ model
has a relaxation time that only diverges at $T=0$ and hence has  
no finite temperature singularity.

In the opposite limit where $K=\infty$ (i.e. on the Bethe lattice), it is 
easily shown that a singularity arises when the probability 
for a site to be unable to relax satisifies the self-consistent 
relation~\cite{MS}: 
\be
P = (1-c) [ P^3 + 3(1-P)P^2 ],
\ee 
which shows that $P(T \le T_c) = P(T_c) + a \sqrt{T-T_c}$ 
and corresponds to the well-known `square root' singularity also found
in the context of mode-coupling theory~\cite{gotze_book} 
and controls for instance the
temperature dependence of the long-time limit of the persistence
function in the glass phase.
 
For intermediate $K$, one can 
readily generalize the $K=1$ argument on the probability. 
The only variant is that 
instead of requiring one vacancy per side of 
an expanding square of size $\ell$~\cite{TBF}, one now requires
at least $K$ consecutive up spins on each side of the square. 
This procedure generates the macro-vacancies
of the KFA model. The probability of such a `$K$-macro-vacancy' 
reads:
\be
P(K) = c^{K^2} \prod_{\ell = K}^\infty [ 1 - (1-c^K)^{\ell/K} ]^4
\sim c^{K^2} \exp(-K c^{-K}) > 0.
\ee  
From this argument, we 
conclude therefore that the KFA model at finite $K$ has no finite $T$
singularity, as for $K=1$, but since the above probability 
$P(K)$ increases rapidly
with $K$, large $K$ values should produce results that are closer
to the mean-field Bethe lattice limit obtained at $K=\infty$. 
The data shown in Fig.~\ref{tau} are 
clearly consistent with this expectation, and confirm also 
that models with larger $K$ have a larger kinetic fragility, i.e.
a sharper temperature dependence. 

\subsection{Four-point dynamic susceptibility}

In this section we discuss the behaviour of the four-point function
in the KFA model, and in particular its behaviour 
across the mode-coupling crossover. 

\begin{figure}
\psfig{file=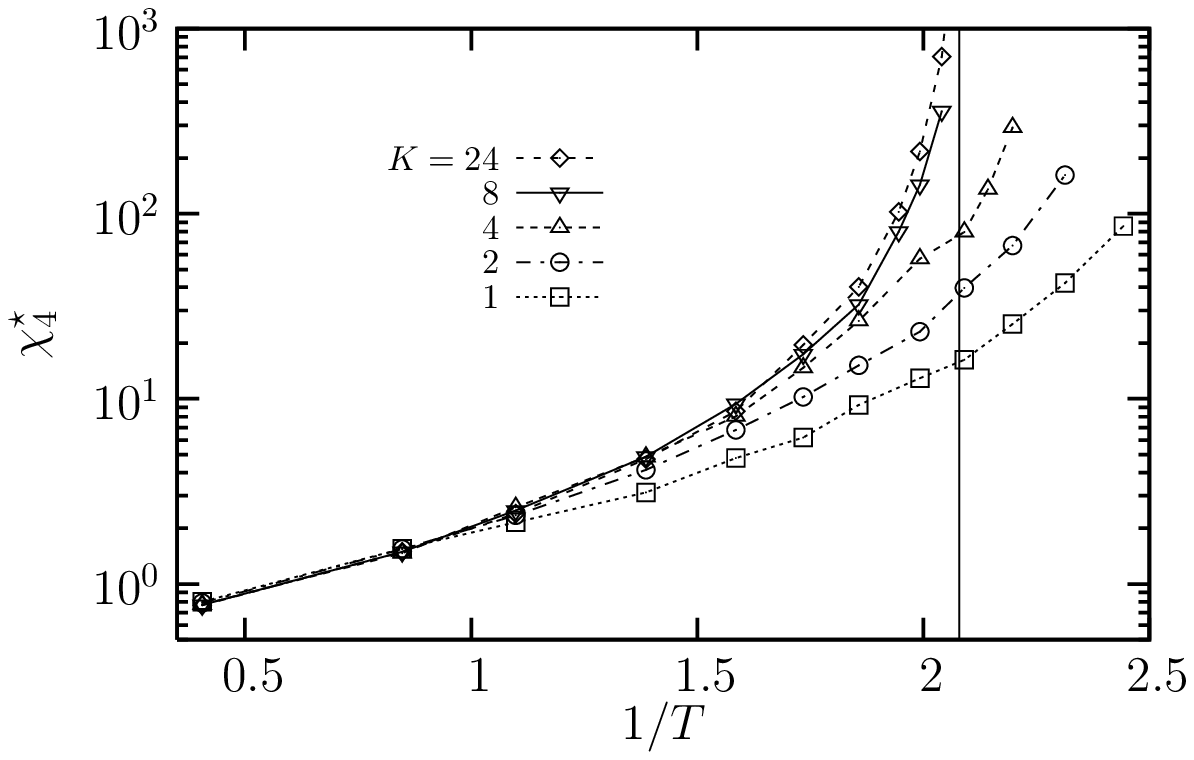,width=8.5cm}
\psfig{file=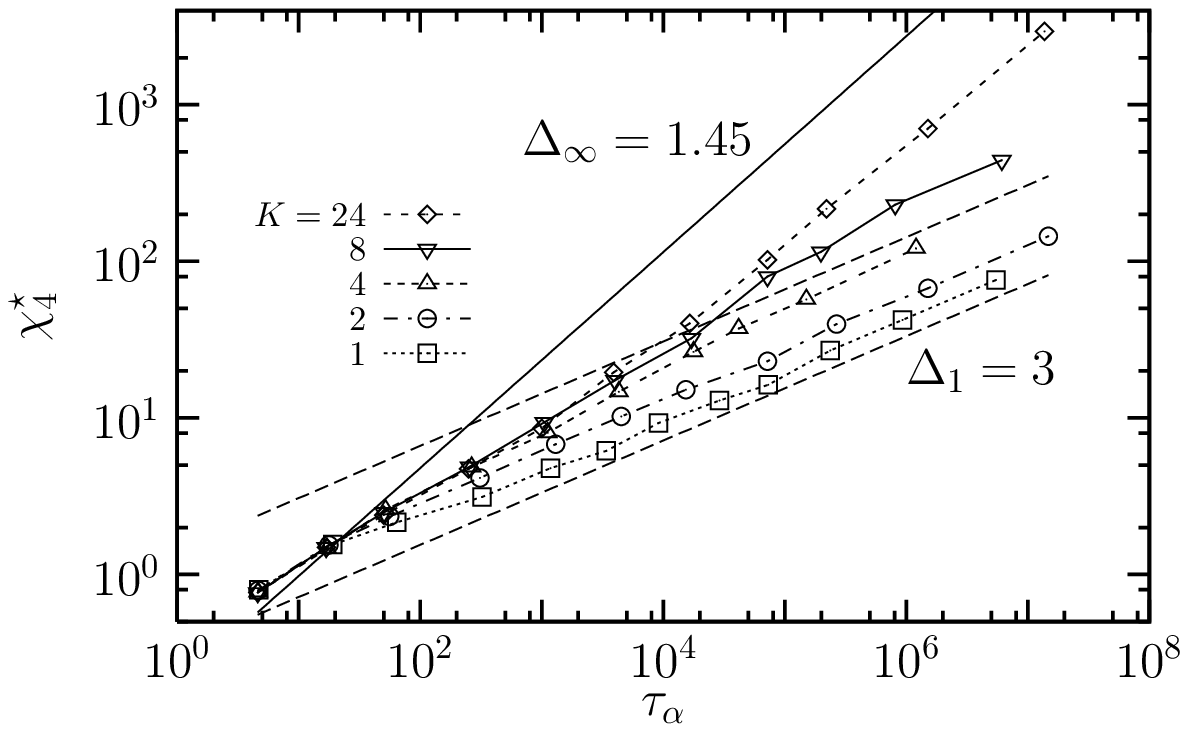,width=8.5cm}
\caption{\label{chis} Top: Temperature evolution of the peak
of $\chi_4$ for the KFA model in an Arrhenius plot. 
Bottom: Dynamic scaling between peak of the susceptibility and 
relaxation time. Power laws, Eq.~(\ref{Delta}), 
with exponents $\Delta_1 = 3$ and $\Delta_\infty= 1.45$ are shown
for comparison.} 
\end{figure}

We define the four-point susceptibility $\chi_4(t)$ in terms of 
the spontaneous fluctuations of the persistence 
function~\cite{steve,nef}: 
\be
\chi_4(t) = N \left[ \langle p^2(t) \rangle - \langle p(t) \rangle^2 \right],
\ee
where $p(t)$ is the instantaneous value of the persistence in a system 
composed of $N$ sites, such that $\langle p(t) \rangle = P(t)$. 
The time evolution of $\chi_4(t)$ is as found in many other systems.
It has a peak whose  
height $\chi_4^\star$ increases when $T$ decreases, and its 
position in time shifts to larger times, essentially 
tracking the alpha-relaxation time. 
As suggested by previous work~\cite{TWBBB}, the approach to this peak 
obeys either a single power for $K=1$, or is composed of 
two distinct power laws in the mode-coupling regime, $K > 2$.  
 
We follow the temperature evolution of the peak of the dynamic
susceptibility, $\chi_4^\star(T)$, in Fig.~\ref{chis}.
We use two representations. In Fig.~\ref{chis}a, we use an Arrhenius 
plot to emphasize the similarity of behaviour of $\chi_4^\star$ with 
the behaviour of $\tau_\alpha(T)$. The crossover nature of these 
curves is in particular very clear, with a near power law divergence for 
$K=24$, or a much slower growth for $K=1$. Interestingly, 
we again find that $\chi_4^\star$ has a mixed behaviour for 
intermediate $K$ values, clearly crossing over from mode-coupling
to cooperative behaviour near $T_c$. 

This crossover becomes more striking when representing 
$\chi_4^\star$ as a function of $\tau_\alpha(T)$ using $T$
as a running parameter, as frequently used in studies
of glass-forming systems~\cite{steve,dalle_07}, see Fig.~\ref{chis}b.
In this representation, both physical regimes
are well-described by a power law `dynamic scaling', 
\be
\chi_4^\star \sim \tau_\alpha^{1/\Delta}, 
\ee 
where the dynamic exponent $\Delta$ for a given $K$ 
takes two values, 
\be 
\Delta(K=\infty) = \Delta_\infty 
\approx 1.45, \quad \quad \Delta(K=1) = \Delta_1 \approx 3.
\label{Delta}
\ee 
For intermediate $K$ values, the data exhibit a clear crossover from 
$\Delta_\infty$ to $\Delta_1$ as $\tau_\alpha$ increases, see for instance 
the data for $K=4$ in Fig.~\ref{chis}b.

It is interesting to note that 
the mode-coupling crossover, in this simple model at least, 
is not characterized by a non-monotonic behaviour of
$\chi_4(t)$, but rather by a change of its temperature dependence. 
Although this reflects nicely the change of physical mechanism 
for structural relaxation near $T_c$, as shown in Fig.~\ref{chis}, 
the behaviour of $\chi_4$ is not as striking as the non-monotonic
size dependence of $\tau_\alpha(L,T)$ in Fig.~\ref{taurfar8b}, 
and it shows no sign of the non-monotonic temperature 
dependence found above for $\ell_{FS}(T)$,
or the non-monotonic dynamic profiles reported near amorphous walls 
in Ref.~\cite{sandalo}.

\end{document}